# Apically Dominant Mechanism for Improving Catalytic Activities of N-Doped Carbon Nanotube Arrays in Rechargeable Zinc-Air Battery

Wenhan Niu[1,#], Srimanta Pakhira[2,3,4,5,#], Kyle Marcus[6], Zhao Li[6], Jose L. Mendoza-Cortes [2,3,4,5,*] and Yang Yang[1,6,*]

[1]NanoScience Technology Center, University of Central Florida, Orlando, FL 32826, United States

[2] Department of Chemical & Biomedical Engineering, Florida A&M University - Florida State University, Joint College of Engineering, Tallahassee, Florida, 32310, USA

[3] Condensed Matter Theory, National High Magnetic Field Laboratory (NHMFL), Florida State University (FSU), Tallahassee, Florida, 32310, USA

[4] Materials Science and Engineering, High Performance Materials Institute (HPMI), Florida State University, Tallahassee, Florida, 32310, USA

[5] Department of Scientific Computing, 400 Dirac Science Library, Florida State University, Tallahassee, FL 32306-4120, USA

[6]Department of Materials Science and Engineering, University of Central Florida, Orlando, FL 32826, United States

[*] *Corresponding author: mendoza@eng.famu.fsu.edu; Yang.Yang@ucf.edu*

[#] *These authors contribute equally to this work.*

# ABSTRACT

The oxygen reduction (ORR) and oxygen evolution reactions (OER) in Zn-air batteries (ZABs) require highly efficient, cost-effective and stable electrocatalysts as replacements to traditionally high cost, inconsistently stable and low poison resistant Platinum group metals (PGM) catalysts. Although, nitrogen-doped carbon nanotube (NCNT) arrays have been developed over recent decades through various advanced technologies are now capable of catalyzing ORR efficiently, their underdeveloped bifunctional property, hydrophobic surface, and detrimental preparation strategy are found to limit practical large-scale commercialization for effective rechargeable ZABs. Here, we have demonstrated fabrication of a three-dimensional (3D) nickel foam supported NCNT arrays with CoNi nanoparticles (NPs) encapsulated within the apical domain (denoted as CoNi@NCNT/NF) that exhibits excellent bifunctional catalytic performance toward both ORR (onset potential of 0.97 V vs. RHE) and OER (overpotential of 1.54 V vs. RHE at 10 mA cm$^{-2}$). We further examined the practicability of this CoNi@NCNT/NF material being used as an air electrode for rechargeable ZAB coin cell and pouch cell systems. The ZAB coin cell showed a peak power density of 108 mW cm$^{-2}$ with an energy density of 845 Wh kg$_{Zn}^{-1}$ and robust rechargeability over 28h under ambient conditions, which exceeds the performance of PGM catalysts and leading non-PGM electrocatalysts. In addition, density functional theory (DFT) calculations revealed that the ORR and OER catalytic performance of the CoNi@NCNT/NF electrode are mainly derived from the d-orbitals from the CoNi NPs encapsulated within the apical dominant end of the NCNTs.

**INTRODUCTION**

Benefiting from high-energy density, rechargeability and environmental benignancy, Zinc-air battery (ZABs) technology has been widely regarded as one of the most promising energy conversion and storage technologies to meet the growing energy demands of large-scale application for electric vehicles and other electricity-related devices.[1-5] The two prominent reactions involved in ZABs are oxygen reduction reaction (ORR) and oxygen evolution reaction (OER), both of which determine the practical performance of ZABs. Thus, the use of highly efficient and stable electrocatalysts as air electrodes is of paramount significance for facilitating the sluggish ORR and OER, thereby attaining high power and long operating lifetime performance for ZABs. Typically, platinum group metals (PGM) are most favorable as ORR and OER catalysts displaying absolute superiority of catalytic activity. However, the high cost, poor stability/durability and low poison resistance of PGM have been the primary barriers that hamper widespread commercialization of ZABs.[2, 6-9]

With decades of extensive research in developing cost-effective catalysts that possess remarkable catalytic performance, a new generation of PGM-free electrocatalysts based on carbon materials such as heteroatom doped carbon nanotubes, graphene, and porous carbons have been studied. Most of these carbon catalysts exhibit comparable activity to PGM catalysts, which can be attributed to the heteroatom-introduced electron density redistribution of surface carbon atoms and an increase in pentagonal graphene edge defects. Among these, nitrogen-doped carbon nanotube (NCNT) arrays have previously shown impressive performance in fuel cells due to their favorable properties such as the in-situ growth of NCNT electrocatalyst on electrode substrate.[5, 10-11] However, many well-developed NCNT arrays present with a hydrophobic surface characteristic that does not allow for $O_2$, $H_2O$, $OH^-$ and other ORR or OER intermediates to sufficiently permeate through the NCNT arrays, creating an isolation of interior active sites. Additionally, the fabrication of carbon nanotube arrays are generally achieved through bottom-up processes by using transitional metal nanoparticles (NPs), mainly distributed at the base of NCNT arrays, as seeds for nanotube growth. This nano-architecture displaying hydrophobic properties at the carbon surface is unfavorable for fully accessible active sites being derived from carbon-encapsulated transition metals at the NCNT array base. Besides, neither the commonly used Si or Cu substrates are suitable for serving as an air electrode in ZABs, due to their air-impenetrability and easy corrosion in alkaline electrolyte.[12-13] Thus, to further improve the performance of carbon nanotube arrays electrocatalysts in fuel cells will be a great challenge using the conventional synthesis strategy.

Inspired by the mechanism of apical dominance in plant growth, where the shoot apex upward growth is crucial for better access to a light source, thus promoting more efficient photosynthesis when compared to lateral bud growth.[14-15] Hence, according to plant physiology, it is conceivable that vertically grown NCNT arrays that present apically dominant active sites are better equipped to absorb $O_2$, $H_2O$, $OH^-$ and other associated reactants for a more immediate ORR or OER response, than compared with the base-grown NCNT arrays. For this reason, it is of great interest to rationally design a NCNT array electrode with enhanced catalytic activity at the NCNT apex. Herein, we have developed a highly efficient PGM-free catalyst comprised of three-dimensional (3D) nickel foam (NF) supported NCNT arrays in which CoNi alloy NPs were rationally implanted into the NCNT apex (the final product denotes CoNi@NCNT/NF). As a result, the as-prepared CoNi@NCNT/NF exhibited excellent catalytic performance toward both ORR and OER. Moreover, the rechargeable ZAB coin cells were assembled using CoNi@NCNT/NF that showed impressive performance in terms of output power density, and long-term cyclability, outperforming equivalent PGM catalysts.

**RESULTS AND DISCUSSION**

The CoNi@NCNT/NF catalyst was synthesized through the following steps (Fig. S1): Firstly, a g-$C_3N_4$ layer acted as the carbon source for NCNTs growth, was polymerized and deposited on nickel foam by one step thermal treatment of melamine-covered nickel foam in air (referred to g-$C_3N_4$@NF, Fig. S1b); Then, $Co_3O_4$ nano-flakes were deposited on the g-$C_3N_4$@NF surface by repeatedly soaking g-$C_3N_4$@NF in cobalt nitrate solution and subsequent heat treatment in a muffle furnace under air atmosphere (denoted as $Co_3O_4$@g-$C_3N_4$@NF, Fig. S1d); Finally, after pyrolysis of $Co_3O_4$@g-$C_3N_4$@NF under nitrogen flow, NCNTs arrays were grown onto the nickel foam by following a "tip-growth" mechanism. More precisely, during heat treatment, the Ni atoms from nickel foam underwent diffusion and subsequent alloying with Co metal to form CoNi alloy NPs that served as catalytic seeds for NCNTs tip-growth. As a result, the NCNTs with CoNi NPs encapsulated within their apical domains were synthesized (denoted as CoNi@NCNT/NF, Fig. S1e).

Scanning electron microscopic (SEM) images in Fig. 1a reveal a 3D structure of the nickel foam with open pathways for electrolyte and gas diffusion. A closer SEM observation in Fig. 1b shows vertical NCNT arrays are homogeneously grown on the surface of nickel foam. The corresponding SEM elemental mapping in Fig. S2 shows that C, N, O, Co and Ni elements uniformly exist within the skeleton of CoNi@NCNT/NF. Interestingly, as depicted in Fig. 1 c, the metal NPs (indicated by yellow circles) are only observed within the apex of each carbon nanotube. To further investigate the growth mechanism of CoNi@NCNT/NF, synthesis of the NCNT/NF without encapsulating CoNi NPs was completed for comparison. Elemental mapping of NCNT/NF indicate that the C, N, O and Ni elements are evenly distributed within the nickel foam surface (Fig. S3), where there is an apparent lack of metal NPs in the carbon nanotube arrays apex, as shown in Fig. S4 a and b. This implies that the growth of the NCNT/NF follows a "base-growth" mechanism, further indicating the incorporation of Co compounds into the fabrication precursor plays a vital role in the apically dominant growth mechanism for CoNi@NCNT/NF.

Transmission electron microscopy (TEM) image (Fig. 1d) of the CoNi@NCNT catalyst further confirms that metal NPs are encapsulated within the carbon nanotube apex. High-resolution transmission electron microscopy (HRTEM) images in Fig. 1e show a well-defined graphite lattice spacing in the range of 0.42-0.54 nm (see Fig. 1f), which is larger than the typical lattice spacing of pristine graphite (0.34 nm). This can be attributed to the addition of nitrogen elemental dopants in the graphite layers, which increases the spacing of carbon layers.[16] This is especially apparent for the carbon layers directly surrounding the metal NPs. The number of carbon layers surrounding the metal NP are also variant and highlighted in Fig. 1e and f as: 1-layer corresponding to the blue and red areas, 3-layer for the yellow area, and 6-layer for the purple area. This implies the metal NP has more effect on the most-outer carbon layer of tip section in NCNT than those of other places. Moreover, a lattice spacing of 0.2 nm is observed for the selected area in HRTEM image (Fig. 1e), which agrees well with the (111) crystal plane of CoNi alloy.[17] TEM elemental mapping shows that N and C signals are uniformly distributed within carbon nanotubes, whereas Co and Ni signals are only detected at the black-contrast apical domains in the upper carbon nanotube sections. These findings suggest that nitrogen element is doped into the carbon nanotubes, and the metal NPs within the carbon nanotubes are comprised of Co and Ni elements. X-ray diffraction (XRD) patterns for both CoNi@NCNT/NF and NCNT/NF show diffraction peaks located at 26° (Fig. S5), which is corresponding to the

(002) crystal plane of CNTs. Notably, the (002) diffraction peak of CoNi@NCNT/NF exhibits higher intensity and a smaller peak width at half height (FWHM) than that of NCNT/NF, indicating the high graphitization degree of CoNi@NCNT/NF.[18] Concurrently, diffraction peaks of the CoNi@NCNT/NF are present at an approximate 0.5º lower than those of nickel metal (JCPDS: 04-0850) and NCNT/NF, indicating the CoNi alloy NPs are encapsulated within the NCNT arrays.

Raman spectra of CoNi@NCNT/NF (Fig. S7) shows two prominent peaks at 1359 and 1587 cm$^{-1}$, corresponding to D and G bands of NCNT, respectively. A higher ratio of D and G band intensities ($I_D/I_G$) indicates a more disordered carbon structure. In our case, the value of $I_D/I_G$ for CoNi@NCNT/NF (0.99) is larger than that of NCNT/NF (0.90), suggesting more defects exist in the carbon layers on the surface near CoNi NPs. X-ray photoelectron spectra (XPS) for both CoNi@NCNT/NF and Ni@NCNT/NF are presented in Fig. S8. As expected, C, N, O and Ni elements are found in both samples, where a Co elemental signal was only detected in the CoNi@NCNT/NF sample. The mole ratio of Co to Ni was estimated to be 0.5 in the CoNi@NCNT/NF. The nitrogen dopants in NCNT can be deconvoluted into pyridinic nitrogen (Pyr-N) and graphitic nitrogen (Gra-N), corresponding to 3.2 at.% for Pyr-N and 5.8 at.% for Gra-N in CoNi@NCNT/NF (Fig. S9). In contrast, there are 4.2 at.% Pyr-N and 5.9 at.% Gra-N (Fig. S10) in NCNT/NF, which are slightly higher than those in CoNi@NCNT/NF. Meanwhile, previous reports suggested that both Pyr-N and Gra-N species can act as active sites for ORR and OER.[19-20] Therefore, it may be expected that a higher catalytic activity would be observed on the NCNT/NF as compared to the CoNi@NCNT/NF. On contrary, the following electrochemical measurements performed in this work have demonstrated that the catalytic activity of CoNi@NCNT/NF was indeed higher than NCNT/NF. We hypothesize that the alloyed CoNi NPs within the CoNi@NCNT/NF may have synergistic effects that further improve nitrogen dopant catalytic activity.[17]

To better understand the effect of apical dominance for improving the catalytic activity of NCNT arrays, we proposed a mechanistic model to illustrate the electrochemical process on CoNi@NCNT/NF and NCNT/NF electrodes, respectively. A typical schematic diagram of the CoNi@NCNT/NF electrode is illustrated in Fig. 2a. We assume that the NCNTs are vertically grown from the NF surface, where much of the electrochemically relevant species have access to only the NCNT array apex, which can be attributed to the hydrophobic NCNT surface preventing molecular permeation. The implantation of CoNi NPs within the NCNT apical domains could promote catalytic efficiency of the surface active sites, thus accelerating ORR/OER processes on the CoNi@NCNT/NF electrode. For comparison, the schematic illustration in Fig. 2b shows a passive catalytic performance for NCNT/NF with scattered reactants ($O_2$, $OH^-$ and $H^+$ etc.) being absorbed on the NCNT surface. These reactants have limited access to the active sites for ORR and OER due to the hydrophobic characteristic of NCNT.

To verify the previous assumptions, ORR and OER catalytic performance for both samples were first evaluated by steady-state linear sweep voltammograms (LSV) in 0.1 M KOH solution. Fig. 2c shows the CoNi@NCNT/NF electrode with an onset potential ($E_{onset}$) of 0.97 V *vs.* reversible hydrogen electrode (RHE) for ORR, which is more positive than those of Pt/C+$RuO_2$/NF ($E_{onset}$ = 0.91 V) and NiCNT/NF ($E_{onset}$ = 0.81 V) electrodes. For OER catalytic performance, the overpotential at 10 mA cm$^{-2}$ is measured to be $E_{10}$ = 1.54 V *vs.* RHE for the CoNi@NCNT/NF electrode, outperforming Pt/C+$RuO_2$/NF ($E_{10}$ = 1.63 V), NCNT/NF ($E_{10}$ = 1.59 V), NF ($E_{10}$ = 1.71 V) electrodes and the other state-of-the-art catalysts (Supplementary Table 1).[21-24] The NCNT arrays also provide the nickel foam with an increased, readily accessible active area, as verified by the high electrochemical double layer capacitance for CoNi@NCNT/NF ($C_{dl}$ = 18.4 mF cm$^{-2}$) and NCNT/NF ($C_{dl}$ = 18.3 mF cm$^{-2}$), which are greater than that of nickel foam ($C_{dl}$ = 2.4 mF cm$^{-2}$, Fig. S11). This suggests that the higher active surface area of NCNT/NF and CoNi@NCNT/NF are a result from the modification of nickel foam with NCNT arrays, leading to effective promotion of electrochemically accessible surface.

Stability is an important parameter for evaluating catalytic performance of electrocatalysts, the chronopotentiometry curves (Fig. 2d) show the stability of the CoNi@NCNT/NF electrode for ORR and OER, which is much higher than those of other electrodes in our study, with almost no decay of operating potential for either ORR or OER after 10h of operation at 10 mA cm$^{-2}$. Interestingly, the OER overpotential for CoNi@NCNT/NF is shown to decrease after stability testing for 10h, which could be attributed to the formation of hydroxylated surface.[25-26] In contrast, an increased overpotential is observed for both the NCNT/NF electrode and the Pt/C+RuO$_2$/NF electrode after ORR/OER testing for 10h (Fig. 2e). Moreover, the potential difference between LSVs of ORR and OER at 10 mA cm$^{-2}$ ($\Delta E_{10}$) on the CoNi@NCNT/NF is decreased by 0.04 V after 10h operation, whereas increased $\Delta E_{10}$ values are observed with the NCNT/NF ($\Delta E_{10}$ = 0.08 V) and Pt/C+RuO$_2$/NF ($\Delta E_{10}$ = 0.05 V) electrodes suggesting outstanding catalytic activity and stability for the CoNi@NCNT/NF electrode in ORR and OER. The same phenomenon can also be obtained through the amperometric (i-t) analysis (Fig. S12).

To gain further insight into the effect of apical dominance on NCNTs, the catalytic contribution derived from the nickel foam should be ruled out. Therefore, the NCNTs from CoNi@NCNT/NF and NCNT/NF electrodes were individually stripped from nickel foam by sonication, and compared by cyclic voltammetry (CV) using rotating disk electrode (RDE) and rotating ring-disk electrode (RRDE) evaluations. As shown in Fig. S13, CV curves are found to exhibit oxygen reduction peaks for all electrodes in $O_2$-saturated KOH solution. The observed oxygen reduction peak for CoNi@NCNT (0.87 V *vs.* RHE) is more positive than those of NCNT (0.71 V *vs.* RHE) and commercial Pt/C (0.83 V *vs.* RHE), suggesting the remarkable electrocatalytic activity of CoNi@NCNT for ORR. LSV curves in Fig. 3a confirm once again the superior ORR catalytic performance of CoNi@NCNT electrode with a positive onset potential ($E_{onset}$) at 0.97 V *vs.* RHE and a half-wave potential ($E_{1/2}$) at 0.87 V *vs.* RHE. In contrast, those values of NCNT electrode ($E_{onset}$ = 0.88 V and $E_{1/2}$ = 0.70 V *vs.* RHE) and Pt/C electrode ($E_{onset}$ = 0.94 V and $E_{1/2}$ = 0.82 V *vs.* RHE) are much smaller than those of CoNi@NCNT, highlighting the enhanced ORR activity of CoNi@NCNT is attributed to the NCNTs being filled with CoNi NPs. In general, the electrical conductivity of electrocatalyst also plays a vital role in ORR performance. The electrochemical impedance spectroscopy (EIS, Fig. S14) of the CoNi@NCNT electrode has a smaller semicircle than that of NCNT electrode, which implies a facilitated ORR process with the CoNi@NCNT electrode arising from the encapsulation of CoNi NPs into NCNTs. Interestingly, the CoNi@NCNT electrode also shows more stable catalytic activity with less $\Delta E_{1/2}$ attenuation (3 mV) than that of Pt/C electrode (32 mV) after 8000 cycles in 0.1 M KOH electrolyte. From the RDE voltammograms in Fig. 3b, it can be observed that the limiting current of CoNi@NCNT electrode increases with an increase in rotation speed (from 225 to 2025 rpm). The corresponding Koutecky-Levich (K-L) plots within the potential range from 0.55 to 0.75 V *vs.* RHE show a good linear relationship and similar slope to that of Pt/C. This suggests first order ORR reaction kinetics, which are related to oxygen concentration in the electrolyte and lead to a four-electron ORR pathway for the CoNi@NCNT electrode. The kinetic current for CoNi@NCNT ($J_k$ = 26.3 mA cm$^{-2}$) is extracted from the fitted K-L plot intercept at 0.65 V *vs.* RHE and is larger than that of Pt/C ($J_k$ = 16.2 mA cm$^{-2}$) and NCNT electrodes ($J_k$ = 9.4 mA cm$^{-2}$) as can be seen in Fig. 3d. This describes a facilitated kinetics process for CoNi@NCNT showing an increase in Gibbs free energy (ΔG) for ORR. Although, nitrogen content in CoNi@NCNT is lower than that of NCNT (Fig. S10), the encapsulation of CoNi NPs into the apical domain of the NCNTs could indeed optimize the accessibility of reactants to the active sites and therefore enhance catalytic activities of surface carbons as reflected by the above experimental observations.

The ORR catalytic performance for CoNi@NCNT were also evaluated by RRDE measurements. The RRDE voltammograms in Fig. 3e show a larger disk current and more positive overpotential of CoNi@NCNT than those of Pt/C. Moreover, the ring current of CoNi@NCNT is as low as that of Pt/C. Concurrently, the electron transfer number (n) and $HO_2^-$ yield for electrochemical reduction of oxygen can be calculated from ring and disk current (Fig. S15). The calculated values of n = 3.94 and $HO_2^-$ yield = 7.8% for CoNi@NCNT are close to n = 3.96 and $HO_2^-$ yield = 2% for the Pt/C electrode, implying a four-electron pathway with high selectivity toward ORR. Despite the CoNi@NCNT electrode performed well in the electro-reduction of oxygen, the actual active sites are still ambiguous since various active species that could be formed on the carbon surface during the pyrolysis, such as N-C, Metal-N/C and metal compounds such as carbides and oxides.[6, 27] To ascertain the identity of the active sites, we studied the electrochemical performance of CoNi@NCNT under different detection experiments. As shown in Fig. 3f, compared with the LSV curve for a fresh CoNi@NCNT electrode (black line), no visible diminishment is detected on the CoNi@NCNT electrode after acid etching (red dashed line). This phenomenon suggests the carbides, oxides and other metal compounds are absent in the active sites of the CoNi@NCNT electrode. Furthermore, for detecting the insoluble Metal-N/C species, 10 mM and 20 mM KSCN were separately added into the electrolyte before LSV scanning. One can see from Fig. 3f, the LSVs for CoNi@NCNT after adding 10 mM and 20 mM KSCN into 0.1 M KOH maintain the same feature to that of the original one, clarifying no Metal-N/C species exist in the active sites of CoNi@NCNT. Moreover, the CoNi@NCNT electrode also shows a strong tolerance to methanol crossover, as demonstrated by unnoticeable changes of current density in LSV curves after the addition of 1 M and 3 M methanol to the electrolyte (Fig. S16a). Whereas an obvious peak at 0.7 V *vs*. RHE appears in the Pt/C electrode LSV in 0.1 M KOH containing 1 M methanol solution, corresponding to the oxidation of methanol. Also, it is not a surprise that the CoNi@NCNT electrode has a superior OER activity than that of NCNT electrode, as evidenced by the LSV curves for OER in Fig. S17. These before mentioned results highlight good activity and stability of CoNi@NCNT ascribe to an optimal balance of surface active site density, nitrogen content and electron conductivity.

In addition, we have computationally investigated the catalytic activity of the CoNi catalyst which was supported or encapsulated by NCNT i.e. CoNi@NCNT catalyst by calculating electronic properties of the CNT, NCNT and CoNi@NCNT systems; see Fig. 4. Our present calculations showed that there is a very small electron density appearing around the Fermi Energy ($E_F$) level of the CNT although two bands overlapped and showed Dirac-point features like graphene,[28] as shown in Fig. 4a. The Dirac point feature was lost in the band structure of the two wall CNT and two bands are just crossed the Fermi level resulting in less electron density around the $E_F$ as depicted in Fig. 4b. This calculation suggests that there is a repulsion between two layers in the two wall CNT, which is a feature that is also found in bilayer graphene.[28] The interesting results were observed in the N-doped single wall CNT (i.e. NCNT), where doping caused a large change in the electronic properties of the NCNT. The present computation found that the bands overlapped below the Fermi level and an electron density appeared around the $E_F$ of the single wall N-doped CNT as shown in Fig. 4c. The structure of the CoNi catalyst with NCNT i.e. CoNi@NCNT catalyst is shown in Figure Xd along with the band structures and density of states (DOSs). The bands are overlapped on each other resulting a large electron density appeared around the $E_F$, which suggests a metallic behavior. To investigate the large electron density which appeared around the $E_F$, we studied contribution of the d-subshell electron density of the CoNi catalyst which was encapsulated by NCNT. The DOSs of the d-subshell electrons of the individuals Co and Ni atoms were computed and plotted in the inset in Fig. 4d. The diagram suggest that the d-subshell electrons of the Ni atoms provide the major electron density around the Fermi level in the total density of states. Therefore, this is consistent with the d-subshell electron density of the Ni atoms plays an important role in improving the catalytic activity of the CoNi@NCNT and thus this catalyst enhances the catalytic performance for overall oxygen reduction reaction (ORR).

To shed light on the ORR catalytic behaviors of, we further investigated the chemical reaction mechanism for the reaction on the surfaces of the CoNi@NCNT. To study the detail ORR mechanism, we computationally designed a cluster model of CoNi cluster and determined the lowest energy state and stability of the CoNi cluster. We found that the CoNi cluster with the spin=3 and multiplicity=4 is the stable conformer and it has the lowest energy state. The CoNi cluster was encapsulated by the N-doped CNT in our model system. To balance the C-valances, extra H-atoms were added in the CoNi@NCNT model system as shown in Fig. 5. We find that ORR starts with O-O chemisorbing onto the surface of CoNi@NCNT. To describe the reaction network for the ORR mechanism using our model system on the surfaces of the CoNi@NCNT, we followed the total four reaction steps below:

(1) First, an $O_2$ molecule is adsorbed on the surface of the CoNi@NCNT catalyst,

$$O_2 + H_2O + e^- + [M] \longrightarrow [M]\text{-OOH} + OH^- \quad (1)$$

The DFT calculations show that $O_2$ prefers to adsorb at the bridge sites on the CoNi@NCNT surface with presence of water. In the above Eqn. 2, M represents the active adsorption site of the CoNi@NCNT. After absorbing the $O_2$ on the surface of the catalyst, it forms [M]-OOH and $OH^-$, and therefore the $4e^-$ mechanisms were included for the sequential reduction reactions of $O_2$, and then reduced by $H_2O$ together with one electron transfer to form [M]-OOH at top sites and a solvated $OH^-$ anion.

(2) Second, the surface of [M] bound [M]-OOH formed in this step gives $OH^-$ as result of a direct one electron reduction and formation of [M]-O

$$[M]\text{-OOH} + e^- \longrightarrow [M]\text{-O} + OH^- \quad (2)$$

(3) In the third step, [M]-O was reacted by one electron reduction with $H_2O$ and formed [M]-OH as an intermediate.

$$[M]\text{-O} + H_2O + e^- \longrightarrow [M]\text{-OH} + OH^- \quad (3)$$

(4) In the last step, [M]-OH was dissociated into [M] and $OH^-$ by absorbing one electron.

$$[M]\text{-OH} + e^- \longrightarrow [M] + OH^- \quad (4)$$

The oxygen evolution reaction pathways are shown in Fig. 5, and the reactions pathways were plotted Gibbs' free energy ($\Delta G$) versus the reaction coordinate. According to our present DFT-D calculations, the effect of water on the binding energies of OH and OOH is more significant than that of $O_2$ and O in agreement with our experiment. The present study showed how variations in binding energies, which is nothing but $\Delta G$, due to water solvation impact the potential energy diagram of the ORR; see Fig. 5. Our results show that water solvation clearly affects the energetics of elementary steps involved in the ORR.

Due to the remarkable catalytic properties toward both ORR and OER have been demonstrated by electrochemical analysis for CoNi@NCNT/NF, it is possible to allow its use in a rechargeable ZAB as an air electrode. For this purpose, a rechargeable ZAB was successfully assembled into a CR2032 coin cell. As illustrated in Fig. 6a, the schematic configuration of the rechargeable ZAB coin cell is comprised of CoNi@NCNT/NF, glass fiber and Zn plate as cathode, separator and anode, respectively. As depicted in Fig. S18, an open-circuit potential (OCP) of 1.40 V was observed at ambient atmosphere by applying a multimeter to the as-constructed ZAB coin cell, suggesting that CoNi@NCNT/NF has excellent catalytic performance even in an open-to-air coin cell configuration. Polarization and power density curves were conducted to evaluate the highest output power of ZAB. As depicted in Fig. 6b, a current density of 146 mA cm$^{-2}$ and a peak power density of 108 mW cm$^{-2}$ are observed for the battery using CoNi@NCNT/NF as cathode, which are higher than 80 mA cm$^{-2}$ and 54 mW cm$^{-2}$ of ZAB coin cell using commercial Pt/C and those of other ZAB tested in coin cell configuration with ambient conditions (Table S2).[2, 19] These results highlight the outstanding catalytic activity of CoNi@NCNT/NF in a ZAB system and are ascribed to the 3D porous structure of nickel foam combined with highly active NCNTs arrays which facilitate $O_2$ diffusion and the optimized surface nature of NCNT arrays by incorporation of CoNi NPs into NCNTs. The specific capacity for our ZAB coin cell was calculated to be 655 mAh $g_{Zn}^{-1}$ at 5 mA cm$^{-2}$ discharging, corresponded to an energy density of 845 Wh $kg_{Zn}^{-1}$. When the discharge current was increased to 20 mA cm$^{-2}$, the specific capacity of our battery reached as high as 631 mAh $g_{Zn}^{-1}$ (energy density of 738 Wh $kg_{Zn}^{-1}$). Note that the energy densities are higher than the state-of-the-art ZAB using N, P co-doped carbon foam catalysts.[2] Meanwhile, the discharging potentials for ZAB using CoNi@NCNT/NF at 5 mA cm$^{-2}$ and 20 mA cm$^{-2}$ were estimated to be over 1.29 V and 1.17 V, respectively. In a contrast, the operating potential of ZAB using commercial Pt/C was only 0.8 V under 20 mA cm$^{-2}$ discharging. These results confirm again a favorable catalytic activity for CoNi@NCNT/NF serving as air electrode in ZAB. Given that the CoNi@NCNT/NF exhibited good bifunctional activity toward both ORR and OER, the discharge and charge polarization measurements for ZAB should also be considered for assessing the rechargeability using CoNi@NCNT/NF electrode. The polarization curves in Fig. 6d reveal that both the discharging and charging curves of the CoNi@NCNT/NF electrode in ZAB show lower overpotentials at different current densities than those of the ZAB comprised of the Pt/C+RuO$_2$/NF electrode. Similarly, the sufficient rechargeability of ZAB coin cell with bifunctional CoNi@NCNT/NF electrode was also identified by the stable galvanostatic charge-discharge profiles in which no obvious potential drop was detected after a long-term cycling of over 28h at 5 mA cm$^{-2}$. It is worthy of note that the difference between discharge and charge potential ($\Delta E_{DC}$) of the ZAB assembled with the CoNi@NCNT/NF electrode ($\Delta E_{DC}$ = 0.52 V) is much smaller than that of the battery using the Pt/C+RuO$_2$/NF electrode ($\Delta E_{DC}$ = 0.8 V) and the leading results reported elsewhere.[2, 5] The rechargeable performance sufficiently proves the highly efficient and stable catalytic activity of the CoNi@NCNT/NF electrode toward both ORR and OER in such a harsh environment of ZAB coin cell. As shown in Fig. S19, it is found that the CoNi@NCNT/NF electrode still maintains a high-density of NCNT arrays on nickel foam surface after 28h cycling test, suggesting robust characteristics with a long-term stability in ZAB.

In a certain sense, a ZAB with high capacity and controllable potentials is desirable for application beyond small electronic devices such hearing aids. For this consideration, we constructed a rechargeable ZAB pouch cell to show that our bifunctional CoNi@NCNT/NF electrode can be applied in a large scale rechargeable ZAB. As shown in Fig. S20a, a rechargeable ZAB pouch cell contains the same configuration to that of the ZAB coin cell. However, the ZAB can be designed with a flexible configuration such as a serial or parallel circuit using the pouch cell. Unsurprisingly, the rechargeable ZAB pouch cell assembled with the CoNi@NCNT/NF electrode can be used to start-up a timer (1.5 V, Fig. S20b). Meanwhile, we also examined the durability of the CoNi@NCNT/NF in the rechargeable ZAB pouch cell by long-term discharge-charge cycling test. As a result, the pouch cell using CoNi@NCNT/NF electrode shows superior cycling performance with almost no attenuation after 810 cycles (18h) to that of the pouch cell using Pt/C+RuO$_2$/NF (756 cycles with 16.8h, Fig. S21). For the application of ZAB with a higher output potential, we also designed a pouch cell with two tandem ZABs, as illustrated in Fig. S22. The as-prepared pouch cell with two tandem ZABs was discharged under 10 mA cm$^{-2}$ over 25h, which exhibited a stable discharge potential at 2.31 V without attenuation (Fig. S23). At the same time, the discharge potentials of 1.06 V and 3.46 V were obtained for the single and the three tandem circuits of ZAB pouch cells, respectively, which suggests an effective strategy to extend the application of our ZAB by controlling of the inner configuration and circuit.

**CONCLUSIONS**

In summary, we have developed a strategy for the fabrication of 3D nickel foam supported NCNT arrays where CoNi NPs were individually encapsulated within the NCNT apex. The as-prepared CoNi@NCNT/NF showed efficient catalytic activity toward both ORR and OER, that can be attributed to synergetic contributions from the carbon-encapsulated CoNi NPs and nitrogen doped carbon sites. As a proof of concept, ZAB coin and pouch cells were assembled using CoNi@NCNT/NF as the air electrode. The developed ZAB exhibited an open-circuit potential of 1.40 V, an energy density of 845 Wh kg$_{Zn}^{-1}$ and a peak power density of 108 mW cm$^-$ [2], as well as excellent durability (over 28h of discharge-charge operation) in an air condition, which are better than those of PGM catalysts and other leading electrocatalysts reported recently. The DFT calculations revealed that the carbon-encapsulated CoNi NPs within the NCNT apical domain, due to the d-electrons, further improved catalytic activity of the outermost nitrogen doped carbon and provided additional catalytically active sites on the electrode surface. Incorporation of the apical dominance theory for NCNT array fabrication will undoubtedly pave a way for rationally designed advanced bifunctional electrodes for renewable energy systems.

## ASSOCIATED CONTENT

**Supporting Information**.

This material is available free of charge via the Internet at http://pubs.acs.org."

Detailed method, material characterizations and electrochemical analysis for morphology, composition, catalytic performances of all samples.

## AUTHOR INFORMATION

**Corresponding Author**

* mendoza@eng.famu.fsu.edu

* Yang.Yang@ucf.edu

**Author Contributions**

Y.Y. and W.N. designed the experiments. W.N. synthesized and characterized the material. S.P. and J.L.M.C. performed and designed the computational and theoretical studies. Z.L. performed the XRD and Raman analysis. Y.Y., W.N. and K.M. analyzed the experimental data, S.P. and J.L.M.C. analyzed the computational simulations. All authors discussed the results and commented on the manuscript.

#These authors contribute equally to this work.

**Notes**

The authors declare no competing financial interests.

**ACKNOWLEDGMENT**

This work was financially supported by the University of Central Florida through a startup grant (No. 20080741). S.P. and J.L.M-C. were supported by Florida State University (FSU). J.L.M-C. gratefully acknowledges the support from the Energy and Materials Initiative at FSU. The authors thank the High Performance Computer cluster at the Research Computing Center (RCC) in FSU for providing computational resources and support.

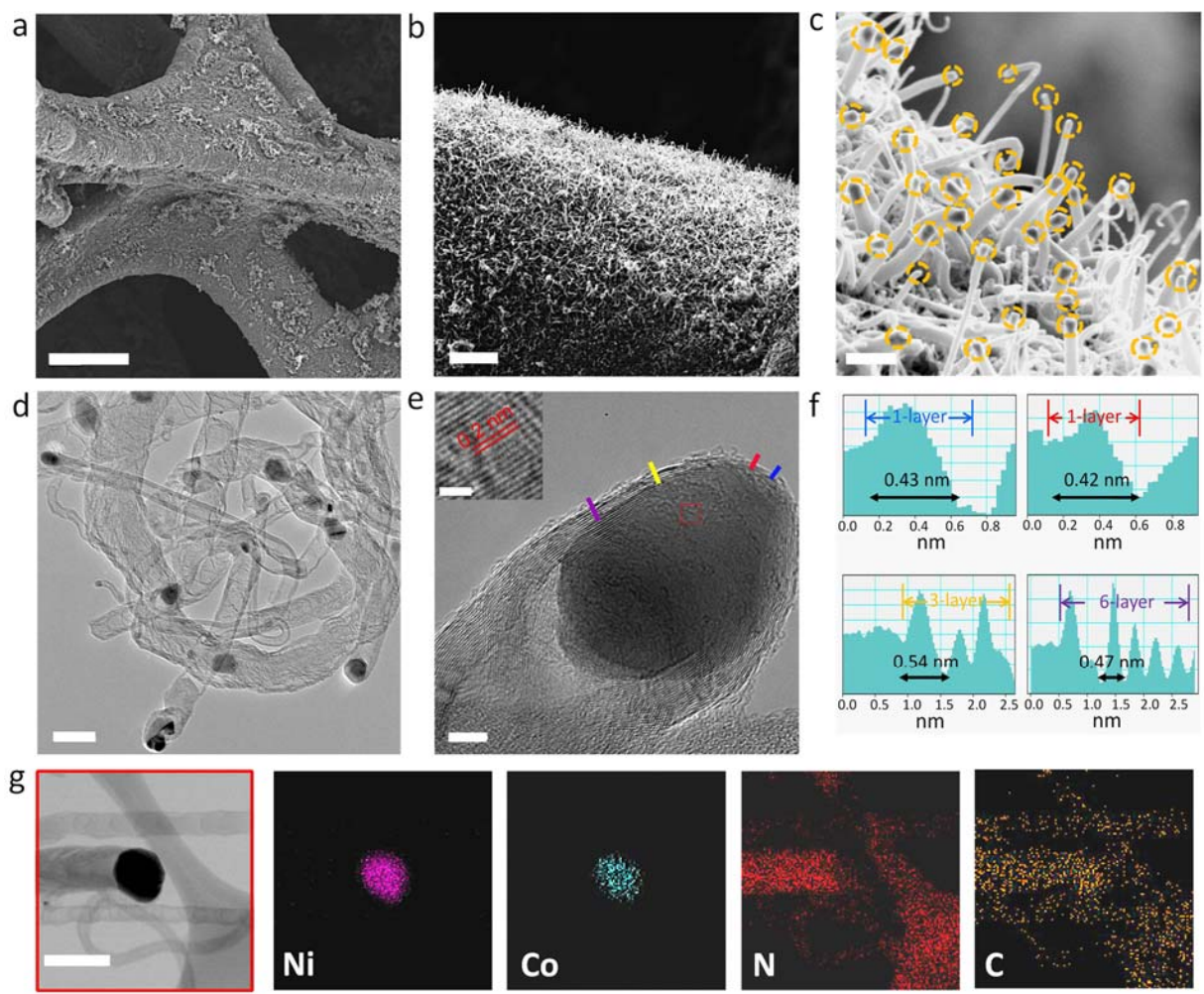

**Figure 1.** (a-c) SEM images, (d) TEM, (e) HRTEM, (f) lattice spacings of graphitic layers corresponding to the marked areas in (e), and (g) elemental mapping of CoNi@NCNT/NF. The yellow cycles marked in (c) indicate the CoNi NPs encapsulated within the top of NCNTs. (e) shows four-colored markers in the graphitic layers, suggesting the blue, red, yellow, purple areas possess 1-, 1-, 3- and 6-layer graphene on the surface of metal NP, respectively. The scale bars in (a-e) and (g) are 20 μm, 5 μm, 200 nm, 50 nm, 5 nm, and 50 nm, respectively. The scale bar in the inset of (e) denotes 1 nm.

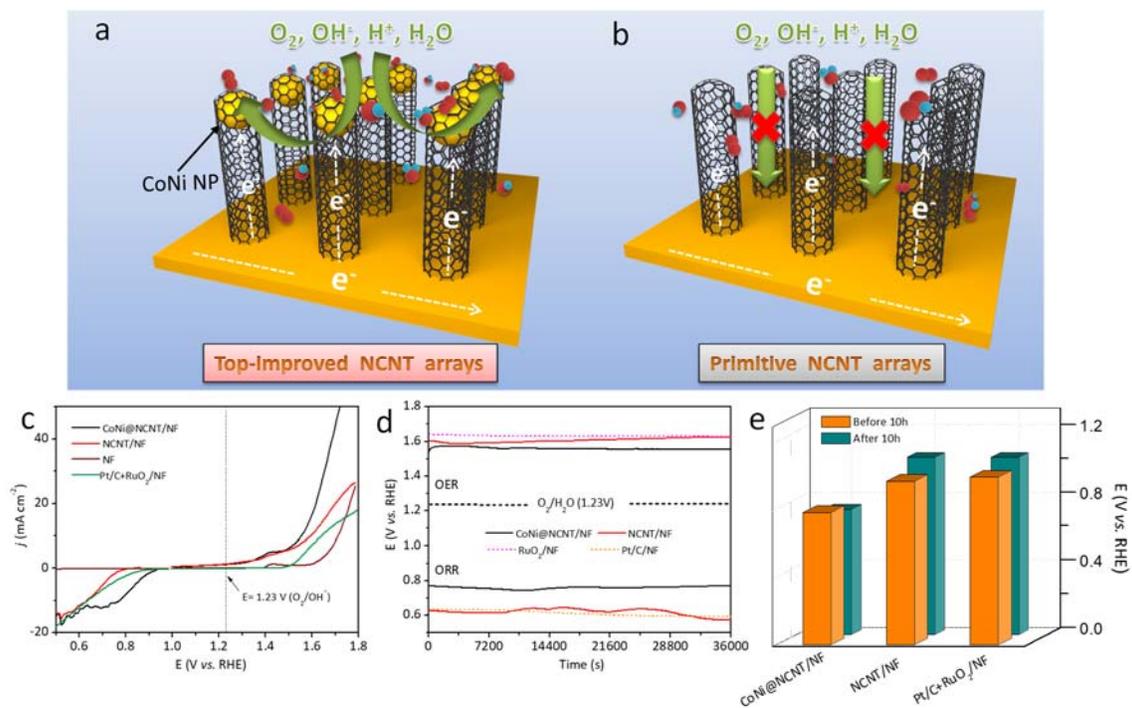

**Figure 2.** Schematic illustrations of (a) CoNi@NCNT/NF and (b) NCNT/NF. (c) LSV curves, (d) chronopotentiometry (v-t) curves and (e) potential differences ($\Delta E_{10}$) between LSV curves at 10 mA cm$^{-2}$ for CoNi@NCNT/NF, NCNT/NF, NF and Pt/C+RuO$_2$/NF electrodes toward ORR and OER in 0.1 M KOH before and after 10h's testing.

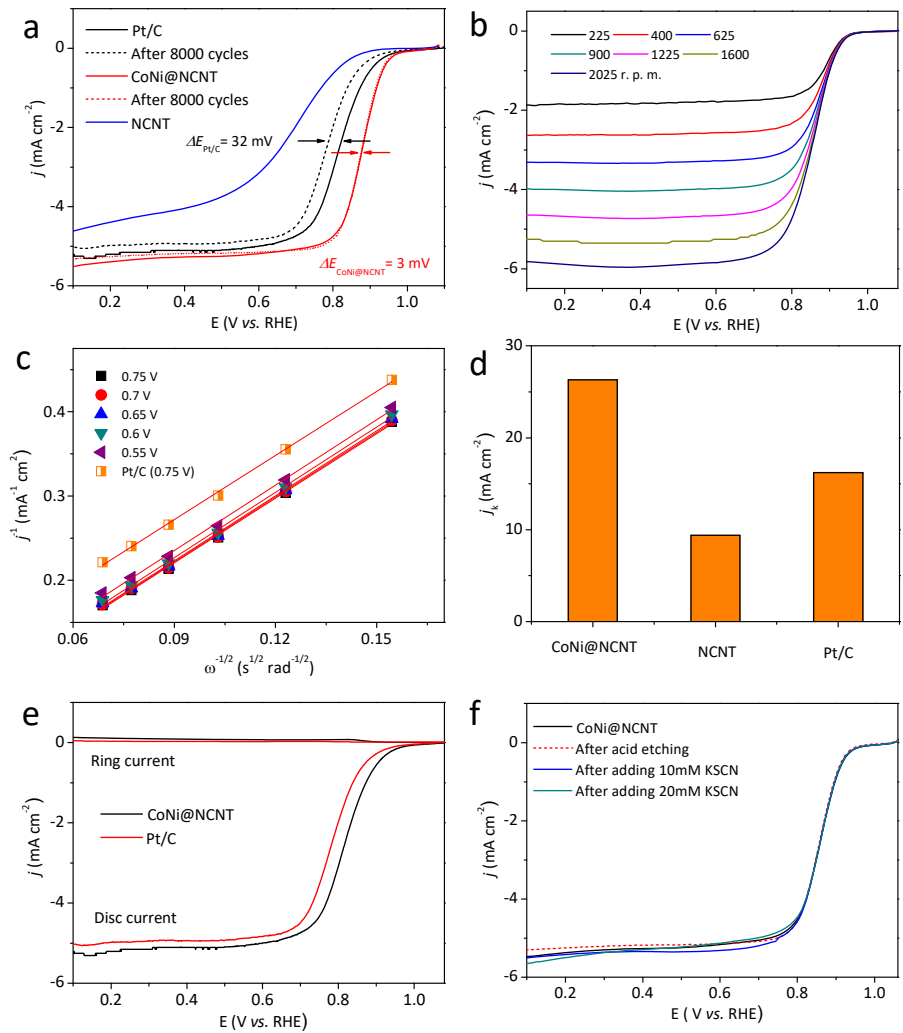

**Figure 3.** (a) Linear sweep voltammogram (LSV) curves for all samples in $O_2$-saturated 0.1 M KOH at 1600 rpm; (b) LSV curves at various rotating speeds and (c) Koutecky-Levich (K-L) plots for the rotating disk electrode (RDE) modified with CoNi@NCNT catalyst; (d) Kinetic current for all samples recorded at 0.7 V *vs.* RHE; (e) Rotating ring-disk electrode (RRDE) voltammograms for CoNi@NCNT and Pt/C in $O_2$-saturated 0.1 M KOH at 1600 rpm; (f) LSV curves of CoNi@NCNT electrode before and after acid etching, and in 0.1 M KOH with 10 mM KSCN and 20 mM KSCN.

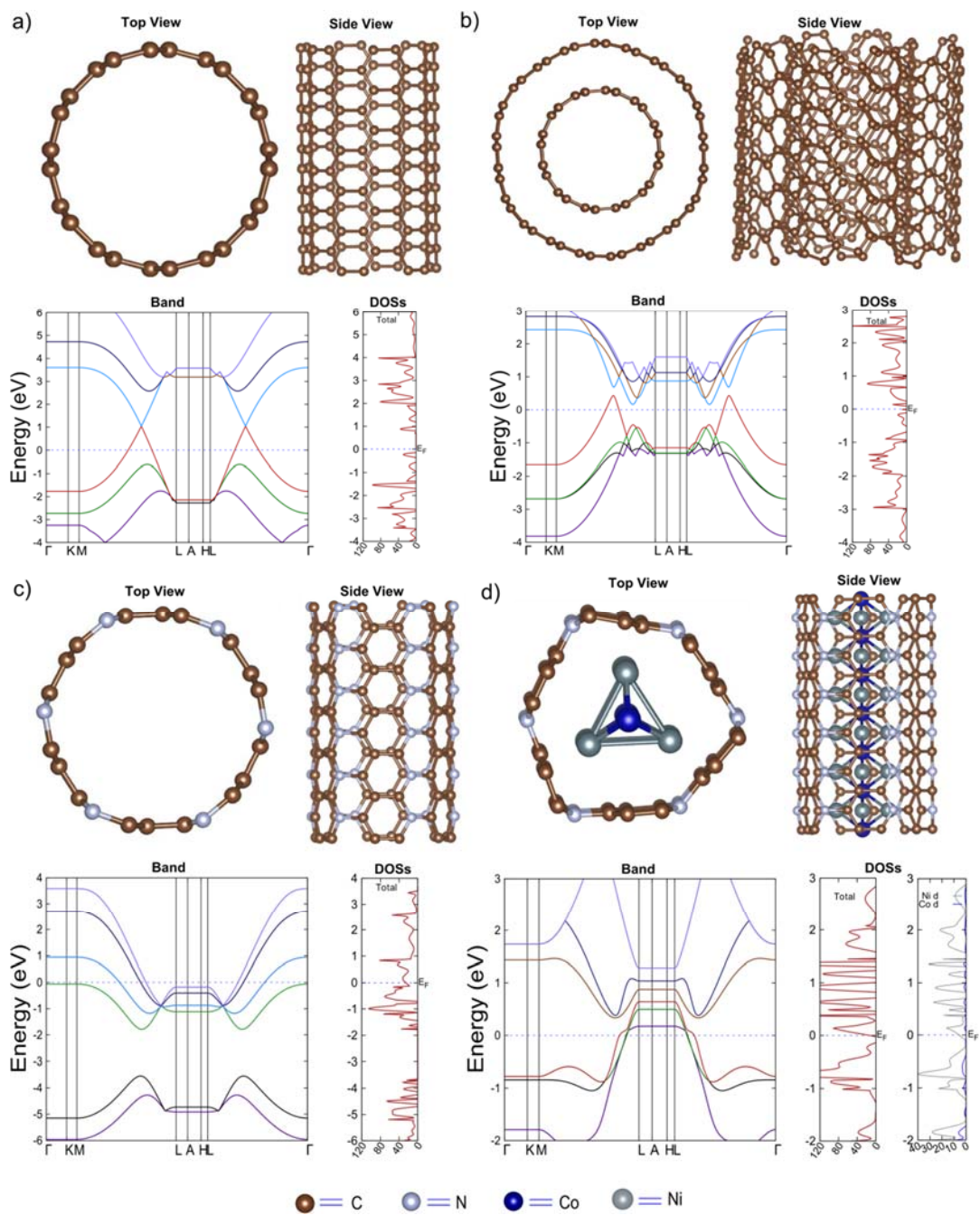

**Figure 4.** Equilibrium structures and electronics properties: band structures and density of states (DOSs) were calculated for (a) single wall CNT; (b) two walls CNT; (c) single wall N-doped CNT (i.e. NCNT); and (d) CoNi catalyst which was encapsulated by NCNT (i.e. CoNi@NCNT). The contribution of the d-subshell electron density of the Ni and Co atoms was shown at the bottom of right hand side of (d).

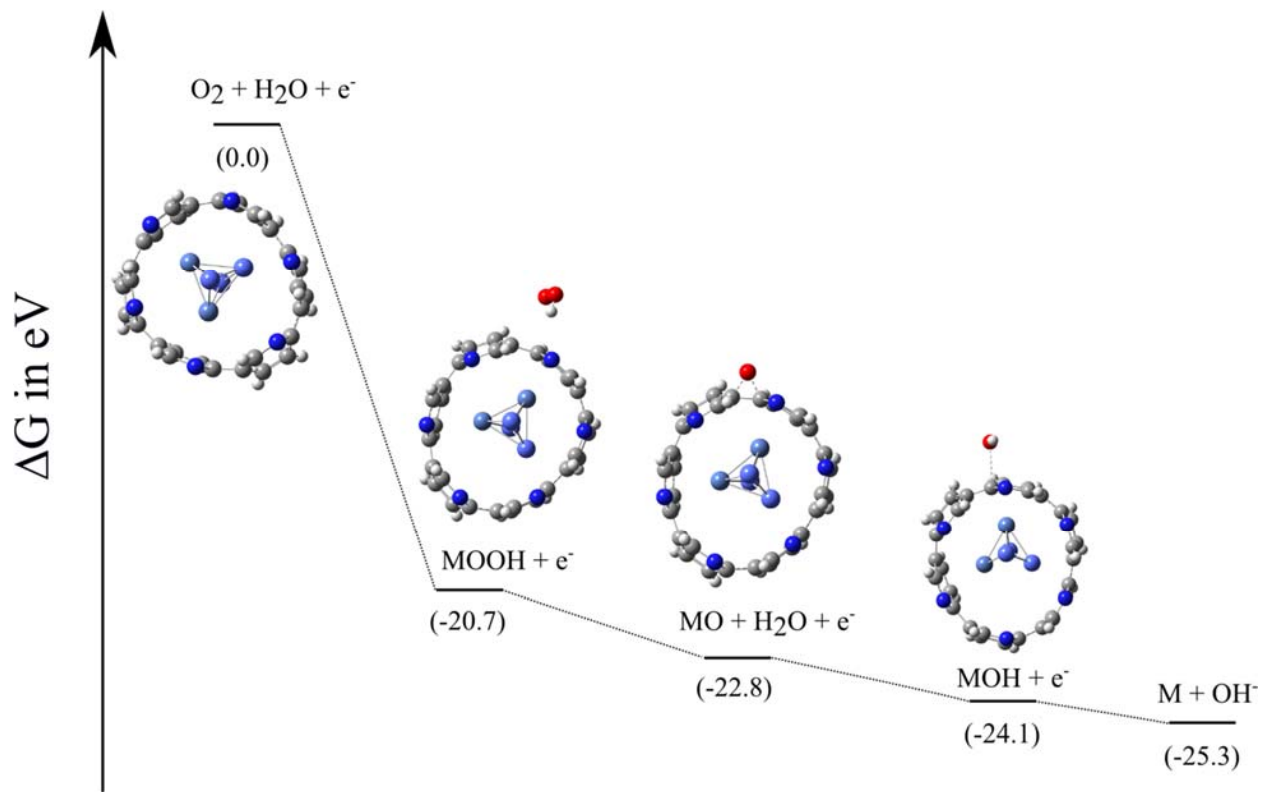

**Figure 5.** Oxygen reduction reaction (ORR) pathways. The relative Gibb's free energy (ΔG) vs. reaction coordinate is shown.

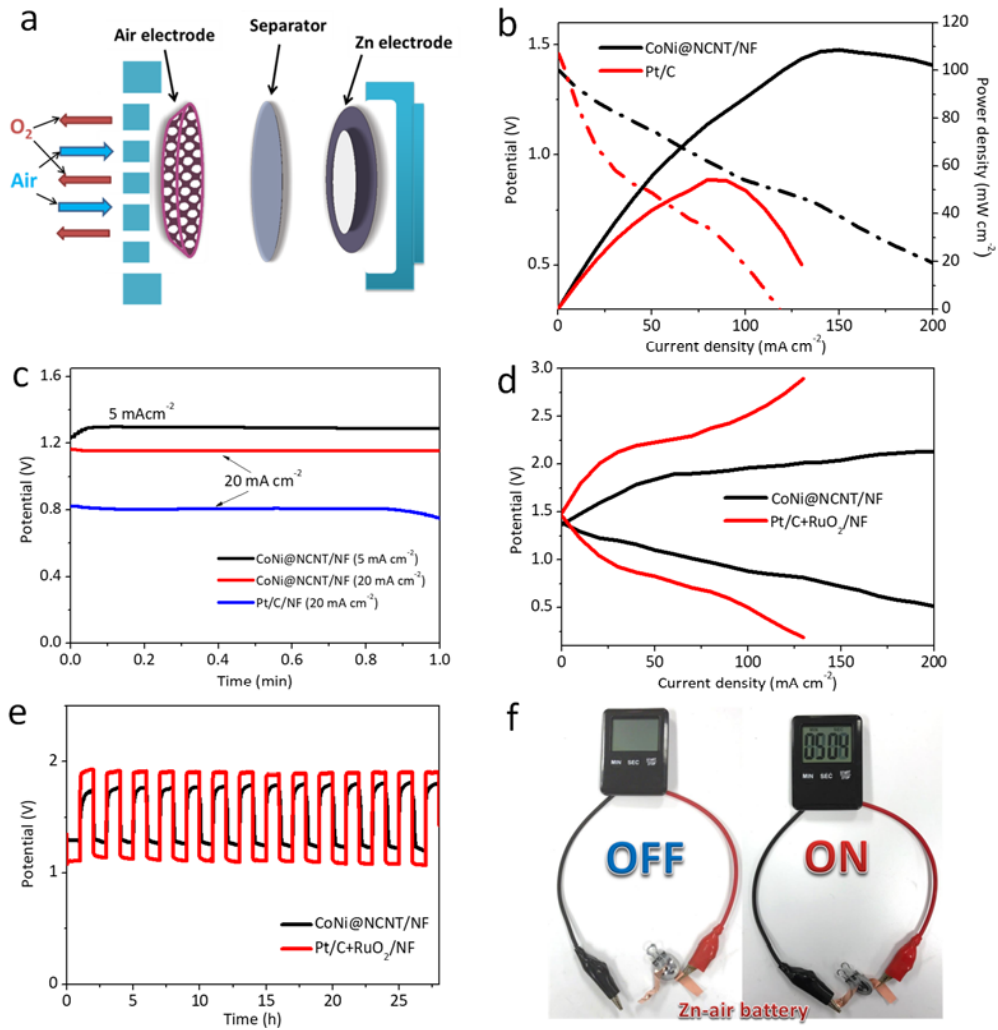

**Figure 6.** (a) Schematic illustration of the typical rechargeable ZAB coin cell configuration. (b) Polarization curves and power density curves of the ZAB coin cells using CoNi@NCNT/NF, NCNT/NF and Pt/C/NF as cathodes. (c) Discharging curves using CoNi@NCNT/NF as cathode at two different current densities compared with the ZAB using Pt/C/NF. (d) Discharging and charging polarization curves, (e) galvanostatic charge-discharge cycling (5 mA cm$^{-2}$, 1 hour for each cycle) of the ZAB coin cells using different air electrodes. (f) Photograph of a timer (1.5 V) driven by the prototype rechargeable ZAB coin cell using CoNi@NCNT/NF cathode.

**TOC**

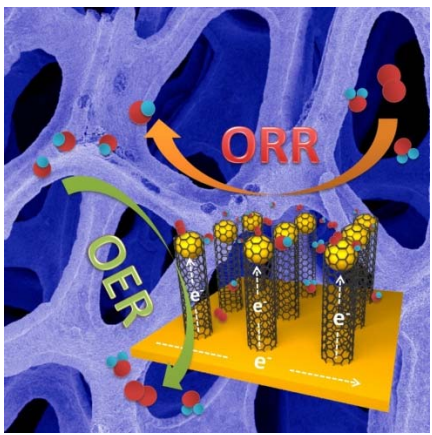

# Apically Dominant Mechanism for Improving Catalytic Activities of N-Doped Carbon Nanotube Arrays in Rechargeable Zinc-Air Battery

Wenhan Niu[1#], Srimanta Pakhira[2,3,4,5#], Kyle Marcus[6], Zhao Li[6], Jose L. Mendoza-Cortes[2,3,4,5*] and Yang Yang[1,6*]

[1] NanoScience Technology Center, University of Central Florida, Orlando, FL 32826, USA

[2] Department of Chemical & Biomedical Engineering, Florida A&M University - Florida State University, Joint College of Engineering, Tallahassee, Florida, 32310, USA

[3] Condensed Matter Theory, National High Magnetic Field Laboratory (NHMFL), Florida State University (FSU), Tallahassee, Florida, 32310, USA

[4] Materials Science and Engineering, High Performance Materials Institute (HPMI), Florida State University, Tallahassee, Florida, 32310, USA

[5] Department of Scientific Computing, 400 Dirac Science Library, Florida State University, Tallahassee, FL 32306-4120, USA

[6] Department of Materials Science and Engineering, University of Central Florida, Orlando, FL 32826, USA

*Corresponding authors: mendoza@eng.famu.fsu.edu (J.L.M-C.) and Yang.Yang@ucf.edu (Y.Y.).

[#] These authors (W.N., S.P.) contribute equally to this work.

# List of Contents

*1. Experimental section*

*2. Characterization*

*3. Supplementary Tables:*

**Table S1.** Comparison of electrocatalytic activities of CoNi@NCNT with leading catalysts reported recently.

**Table S2.** The performance of reported solid or quasi-solid rechargeable Zn-air battery and our Zn-air coin battery.

*4. Supplementary Figures:*

**Fig. S1.** Schematic illustration of the synthesis process for CoNi@NCNT/NF and NCNT/NF.

**Fig. S2.** SEM image and elemental mapping of CoNi@NCNT/NF.

**Fig. S3.** SEM image and elemental mapping of NCNT/NF

**Fig. S4.** SEM images of (a, b) NCNT/NF and (c, d) CoNi@NCNT/NF.

**Fig. S5.** XRD patterns of CoNi@NCNT/NF and NCNT/NF.

**Fig. S6.** (a) BET surface area and (b) pore size distribution of CoNi@NCNT/NF

**Fig. S7.** Raman spectrums of CoNi@NCNT/NF, Ni@NCNT/NF and CoNi@NCNT/NF after 28h testing in Zn-air battery.

**Fig. S8.** XPS survey spectra of CoNi@NCNT/NF and NCNT/NF.

**Fig. S9.** High resolution XPS spectra of N 1s for CoNi@NCNT/NF.

**Fig. S10.** High resolution XPS spectra of N 1s for NCNT/NF.

**Fig. S11.** CVs of (a) CoNi@NCNT/NF, (b) NCNT/NF and (c) NF at various scan rates and the corresponding linear fitting of the capacitive current densities against the scan rates.

**Fig. S12.** Amperometric (i-t) curves for CoNi@NCNT/NF, NCNT/NF, NF and RuO$_2$/NF electrodes at 1.6 V *vs*. RHE.

**Fig. S13.** CVs, LSV curves and K-L plots for CoNi@NCNT, NCNT and Pt/C samples in N$_2$ and O$_2$ saturated 0.1 M KOH.

**Fig. S14.** Electrochemical impedance spectra of CoNi@NCNT and NCNT recorded in O$_2$-saturated 0.1 M KOH.

**Fig. S15.** Electron transfer number (n) and HO$_2^-$ yield ratio of all samples.

**Fig. S16.** (a) LSVs of CoNi@NCNT electrode before and after adding 1 M or 3 M methanol into O$_2$-saturated 0.1 M KOH. (b) LSVs of Pt/C before and after adding 1 M methanol into O$_2$-

saturated 0.1 M KOH.

**Fig. S17.** LSVs of CoNi@NCNT and NCNT electrodes in $O_2$-saturated 0.1 M KOH.

**Fig. S18.** Optical image of an open circuit voltage with 1.4 V for the CoNi@NCNT/NF assembled in a rechargeable Zn-air button battery.

**Fig. S19.** SEM images of the CoNi@NCNT/NF electrode after the long-term galvanostatic charge-discharge cycling over 28h.

**Fig. S20.** (a) Schematic illustration of a rechargeable Zn-air pouch cell using CoNi@NCNT/NF electrode. (b) Optical images of a timer (1.5 V) being driving by the rechargeable pouch cell Zn-air battery using CoNi@NCNT/NF as cathode.

**Fig. S21.** Galvanostatic charge-discharge cycling of the rechargeable Zn-air battery pouch cell using CoNi@NCNT/NF cathode compared with the battery using Pt/C+$RuO_2$/NF cathode.

**Fig. S22.** Schematic illustration of two tandem Zn-air circuits in pouch cell using CoNi@NCNT/NF electrode.

**Fig. S23.** Discharge curves of the Zn-air pouch cell with one single, two tandem and three tandem Zn-air circuits using CoNi@NCNT/NF cathode.

**Experimental section**

*Synthesis of CoNi@NCNT/NF*

The synthesis of CoNi@NCNT/NF contains the following three steps: i) Firstly, one piece of cleaned nickel foam (denotes as NF, the size is 3.5 cm length and 2 cm width) was fully coated with 5 g melamine powder in a ceramic boat by repeated compression, and followed by heat treating of the melamine-coated nickel foam in a furnace with air atmosphere within 1h to form g-$C_3N_4$-coated nickel foam (as depicted in Fig. S1b, the product denotes $C_3N_4$/NF); ii) Then, the $C_3N_4$/NF was immersed in 0.1 M cobalt nitrate solution, dried and subsequently heat treated in furnace within 10-20s to allow the $C_3O_4$ deposition on $C_3N_4$/NF surface. This procedure was repeated three times to form $Co_3O_4$/$C_3N_4$/NF (Fig. S1d); iii) Finally, the CoNi@NCNT/NF was obtained by heat treating of $Co_3O_4$/$C_3N_4$/NF in CVD at 700 ºC with $N_2$ flow followed by washing and drying.

*Synthesis of NCNT/NF*

The NCNT/NF was synthesized following the same procedure to the CoNi@NCNT/NF, except that the second step of $Co_3O_4$ deposition was removed.

**Characterization**

The samples were characterized by a field-emission scanning electron microscope (FE-SEM, ZEISS ultra 55) and a high-resolution JEOL JEM-2100F TEM at 200 kV. Raman spectra were obtained using a Renishaw Raman RE01 scope (Renishaw, Inc.) with a 532 nm excitation argon laser. The chemical state of elements were detected by X-ray photoelectron spectra (XPS) measurements using a PHI Quantera SXM scanning X-ray microscope. The X-ray diffraction (XRD) patterns were obtained by using a powder XRD system (Rigaku D/Max Ultima II, Cu Kα radiation). Nitrogen adsorption-desorption isotherms were measured at 77 K by a Quantachrome Autosorb-3b Brunauer-Emmett-Teller (BET) Surface Analyzer. Before measurements, all samples were degassed at 383 K under vacuum for 2h.

*Electrochemical analysis*

All electrochemical measurements were conducted with a CHI 760E electrochemical workstation in a conventional three-electrode cell, with platinum foil as the counter electrode, Ag/AgCl as the reference electrode, and a catalyst-modified glassy carbon electrode (GCE) as the working electrode. The catalyst inks were fabricated by dispersing 2 mg of catalyst into a solution containing deionized water, isopropanol and Nafion (5%) at a volume ratio of 4:1:0.025 to form a homogeneous suspension with the catalyst concentration being 2 mg mL$^{-1}$. A calculated amount of suspension was then evenly casted on a cleaned GCE surface via syringe and dried in air. Catalyst loading was calculated to be 408 μg cm$^{-2}$ for as-prepared catalysts, 204 μg cm$^{-2}$ for Pt/C and RuO$_2$, respectively. For steady-state linear sweep voltammetry (LSV) measurements, the working electrode was prepared by tailoring CoNi@NCNT/NF, NCNT/NF into 1 x 1 cm$^2$ area. The Pt/C+RuO$_2$/NF electrode was prepared by drop casting the catalyst ink with 2 mg Pt/C and 1 mg RuO$_2$ into nickel foam, and subsequently drying in air.

LSVs for ORR were acquired in an O$_2$-saturated 0.1 M KOH aqueous solution at various rotating speeds (225-2025 rpm) with a scan rate of 5 mV s$^{-1}$. CVs of all samples were tested in a potential range of 0 to -1 V at a scan rate of 50 mV s$^{-1}$. Electrochemical impedance spectroscopy (EIS) measurements were performed by applying an AC voltage with 5 mV amplitude in a frequency range from 100000 to 1 Hz and recorded at 0.85 V (E *vs.* RHE). The electrical double layer capacitance (C$_{dl}$) of the all catalysts were measured from double-layer charging curves using cyclic voltammograms (CVs) in a potential range of 1.08-1.18 V. Before collecting the CV data, the working electrodes were scanned for several cycles until their voltammograms were stabilized. The plot of the current density against scan rate was collected at 1.16 V, which shows a linear relationship, with its slope as the double layer capacitance (C$_{dl}$).

The number of electron transfer (n, eq. 1) and the HO$_2^-$ yield (eq. 2) in oxygen reduction were estimated by the following equations:

$$n = \frac{4I_{Disk}}{I_{Ring}/N + I_{Disk}} \quad (1)$$

$$HO_2^- = \frac{200 I_{Ring}/N}{I_{Ring}/N + I_{Disk}} \quad (2)$$

where N is the collection efficiency (37%), I$_{Disk}$ and I$_{Ring}$ are the voltammetric currents at the disk and ring electrodes, respectively.

The rechargeable Zn-air coin batteries (CR2032 coin cell) tests were performed in a two electrode cell with CoNi@NCNT/NF or Ni@NCNT as the cathode (For precious metal electrocatalysts, the Pt/C or RuO$_2$ ink within ethanol/Nafion solution was drop-casted on nickel foam electrode as air electrode), a polished Zn plate as anode and a glass fiber containing 6 M KOH and 0.2 M Zn(OAc)$_2$·6H$_2$O as the separator. The battery testing was performed in ambient environment on a LAND CT2001A instrument.

*Computational methods*

*Periodic Calculations.*

We have studied the equilibrium structures and electronic properties of the CNT, NCNT and CoNi@NCNT using periodic hybrid dispersion-corrected density functional theory; B3LYP-D3,[1-4] or DFT-D for short, as implemented in CRYSTAL17.[5] In the present computations, we have incorporated van der Waals (vdW) "-D3" dispersion corrections proposed by Grimme, which includes an additional $R^{-8}$ term in the dispersion series.[2-4] Triple-zeta valence with polarization function quality (TZVP) basis sets were used for the C, N, Co and Ni atoms in the periodic DFT-D calculations.[6] The threshold used for evaluating the convergence of the energy, forces, and electron density was $10^{-7}$ (a.u.) for each parameter. Integrations inside of the first Brillouin zone were sampled on 4 x 4 x 8 Monkhorst-Pack[7] k-mesh grids for all the CNT, NCNT and CoNi@NCNT systems during geometry optimization and 20 x 20 x 20 Monkhorst-Pack k-mesh grids for the calculations of band structure and density of states.

*Non-periodic Calculations.*

We have also performed a hybrid dispersion-corrected DFT-D (i.e. B3LYP-D3) calculation to investigate the catalytic activity of the CoNi@NCNT considering a molecular cluster model system.[1-4] DFT-D calculations of all the molecular systems studied here were performed using the general purpose electronic structure quantum chemistry program Gaussian 09 package suite with the default convergence criteria.[8] The 6-31+G** Gaussian type basis sets were used for the H, C, N and O atoms[9] as well as LANL2DZ with effective core potentials (ECPs) were also used for both the Co and Ni atoms.[10-11] The LANL2DZ basis set with ECPs, combines the efficiency of a core-potential-containing basis set with the accuracy of all-electron basis sets such as 6-31+G**, and is well balanced with such basis sets. The harmonic vibrational analysis was performed at the optimized geometry using the same level of theory to obtain the zero-point vibrational energy (ZPE). The vibration calculations did not have any imaginary frequencies, that confirms the stability of the present compounds. The reaction enthalpy changes at 298 K was calculated as the sum of the changes in the electronic energy and the calculated enthalpy corrections. The DFT-D method was used for geometry optimization because densities and energies obtained with the method are less affected by spin contamination than other approaches.[12-14]

The corresponding relative Gibb's free energy of the ORR was calculated as

$$\Delta G = \Delta E + T\Delta S_{vib} + \Delta ZPE + \Delta H_{vib} \qquad (3)$$

where ΔE is the electronic energy and TΔS$_{vib}$ represented the entropic contributions. Zero-point energy (ZPE) corrections were also included based on the calculated vibrational frequencies. We computed entropy contributions to the energies at ambient temperature via frequency analysis, and used the enthalpy (ΔH$_{vib}$) and entropy (ΔS$_{vib}$) outputs to calculate free energies of formation for all relevant intermediates using the standard established protocol.

**Supplementary Table 1.** Comparison of electrocatalytic activities of CoNi@NCNT/NF with leading catalysts reported recently.

| Catalyst | OER $E_{j=10}$ (V vs RHE) | ORR $E_{onset\ potential}$ (V vs RHE) | $\Delta E = E_{j=10} - E_{onset\ potential}$ (V) | Catalyst loading (mg/cm$^2$) | Electrolyte | Ref. |
|---|---|---|---|---|---|---|
| CoNi@NCNT/NF | 1.54 | 0.97 | 0.57 | ~8.45 | 0.1M KOH | This work |
| Co−C$_3$N$_4$/CNT | 1.61 | ~0.90 | ~0.72 | 0.80 | 0.1 M KOH | J. Am. Chem. Soc., 2017, 139, 3336. |
| Co$_3$O$_4$-NCNT/SS | 1.55 | 0.95 | ~0.60 | NA | 0.1 M KOH | Adv. Mater. 2016, 28, 6421 |
| g-C$_3$N$_4$ and Ti$_3$C$_2$ nanosheets (TCCN) | 1.65 | ~0.80 | ~0.82 | 0.21 | 0.1 M KOH | Angew. Chem. Int. Ed. 2016, 55, 1138 |
| NiCo/PFC | 1.63 | 0.92 | 0.71 | 0.13 | 0.1 M KOH | Nano Lett. 2016, 16, 6516 |
| Co@Co$_3$O$_4$/NC-1 | 1.65 | 0.80 | 0.85 | 0.21 | 0.1 M KOH | Angew. Chem. Int. Ed. 2016, 55, 4087 |
| NPMC-1000 | >1.75 | 0.95 | >0.8 | 0.15 | 0.1 M KOH | Nat. Nanotechnol. 2015, 10, 444 |
| phosphorus-doped g-C$_3$N$_4$/carbon-fiber paper | 1.63 | ~0.84 | 0.79 | ~0.20 | 0.1 M KOH | Angew. Chem. Int. Ed. 2015, 54, 4646 |
| H-Pt/CaMnO$_3$ | 1.80 | ~0.95 | ~0.85 | 0.085 | 0.1 M KOH | Adv. Mater. 2014, 26, 2047 |
| Mn$_x$O$_y$/N-doped carbon | 1.68 | 0.81 | 0.87 | 0.21 | 0.1 M KOH | Angew. Chem. Int. Ed. 2014, 53, 8508 |
| S/N_Fe-27 | 1.78 | 0.95 | 0.83 | 0.80 | 0.1 M KOH | J. Am. Chem. Soc. 2014, 136, 14486 |
| N-graphene/CNT | 1.65 | >0.69 | >0.96 | 0.20 | 0.1 M KOH | Angew. Chem. Int. Ed. 2014, 53, 6496 |

**Supplementary Table 2.** The performance of reported solid or quasi-solid rechargeable Zn-air battery and our Zn-air coin battery.

| Catalyst | Peak power density (mW cm$^{-2}$) | Discharge potential (V) | Charge potential (V) | Ref. |
|---|---|---|---|---|
| CoNi@NCNT/NF | ~108 | 1.29 V | 1.81 V | This work |
| NPMC-1000 | 55 | ~0.9 V | ~2.7 V | Nat. Nanotechnol. 2015, 10, 444 |
| CoO$_{0.87}$S$_{0.13}$/GN | ~100 | ~1.1 V | ~1.9 V | Adv. Mater. 2017, 1702526 |
| NCNT/CoO-NiO-NiCo | NA | ~1.18 V | ~2.0 V | Angew. Chem. Int. Ed. 2015, 54, 9654 |
| CNT sheet | NA | ~0.9 V | ~2.0 V | Angew. Chem. Int. Ed. 2015, 54, 15390 |
| Co$_3$O$_4$/NPGC | NA | ~1.18 V | ~1.9 V | Angew. Chem. Int. Ed. 2016, 55, 4977 |

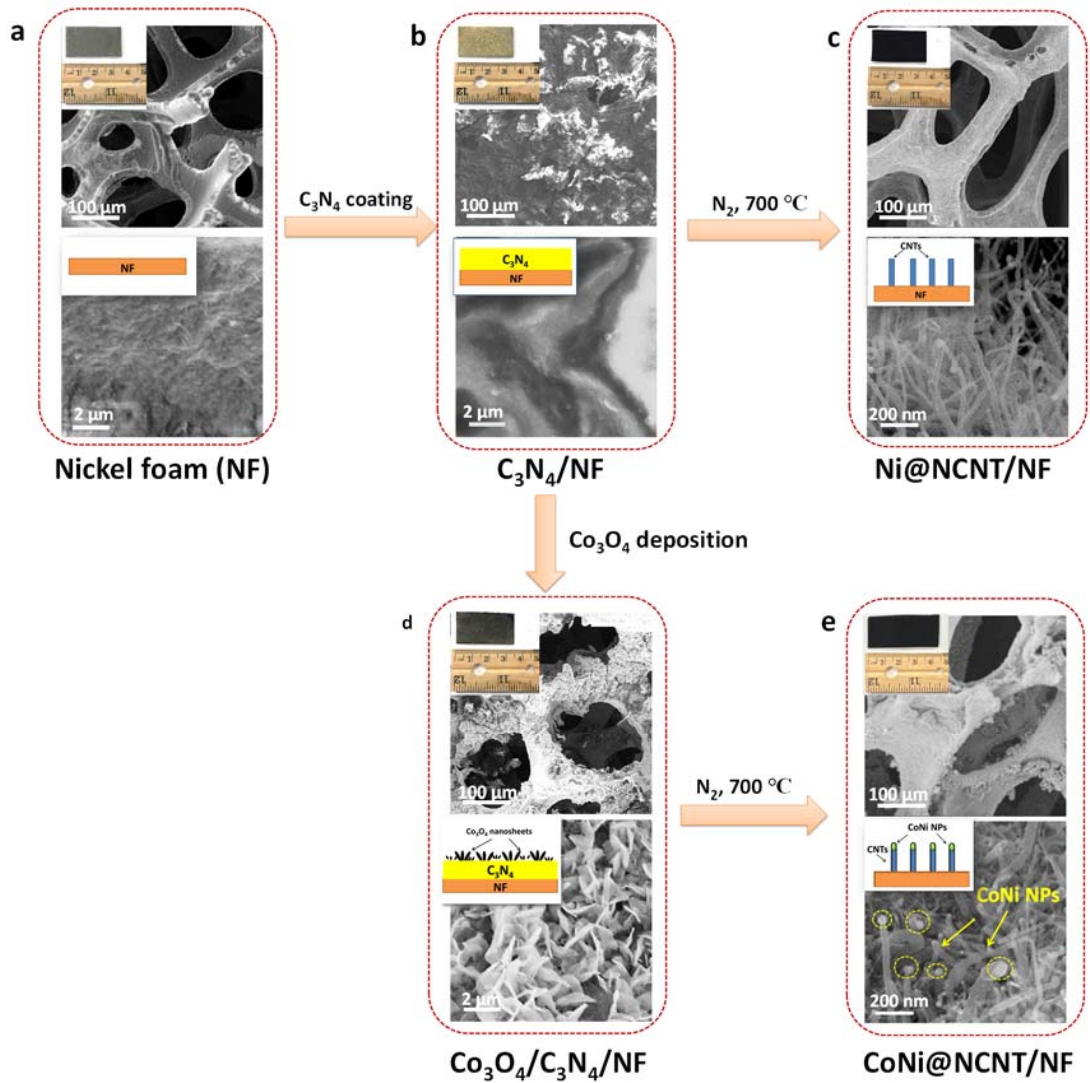

**Fig. S1.** Schematic illustration of the synthesis process for CoNi@NCNT/NF and NCNT/NF.

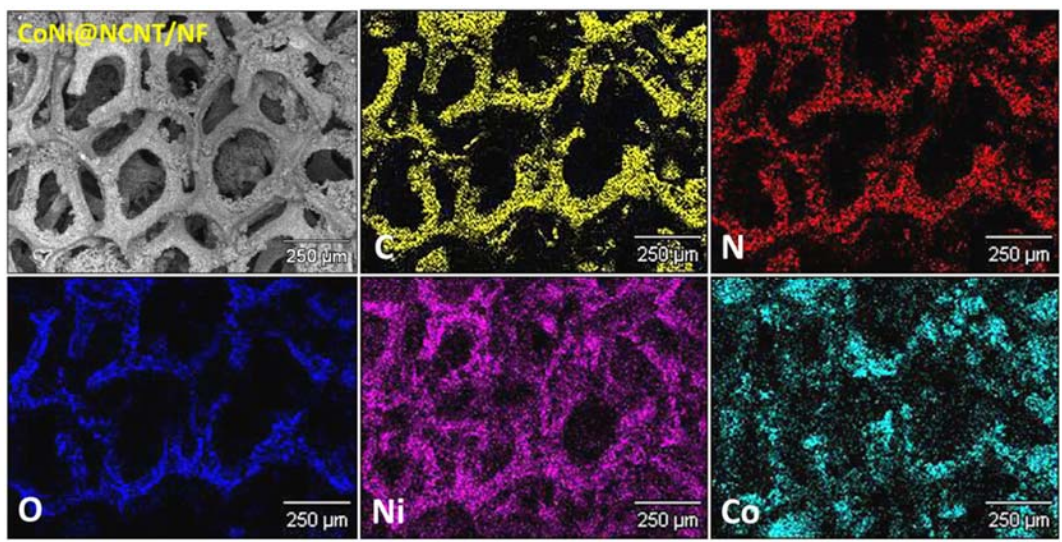

**Fig. S2.** SEM image and elemental mapping of CoNi@NCNT/NF

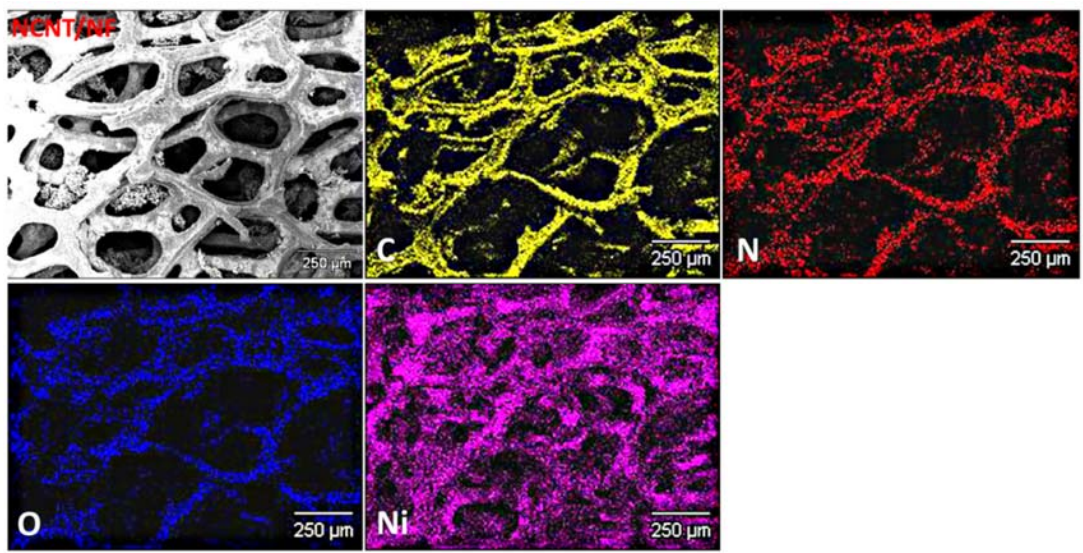

**Fig. S3.** SEM image and elemental mapping of NCNT/NF

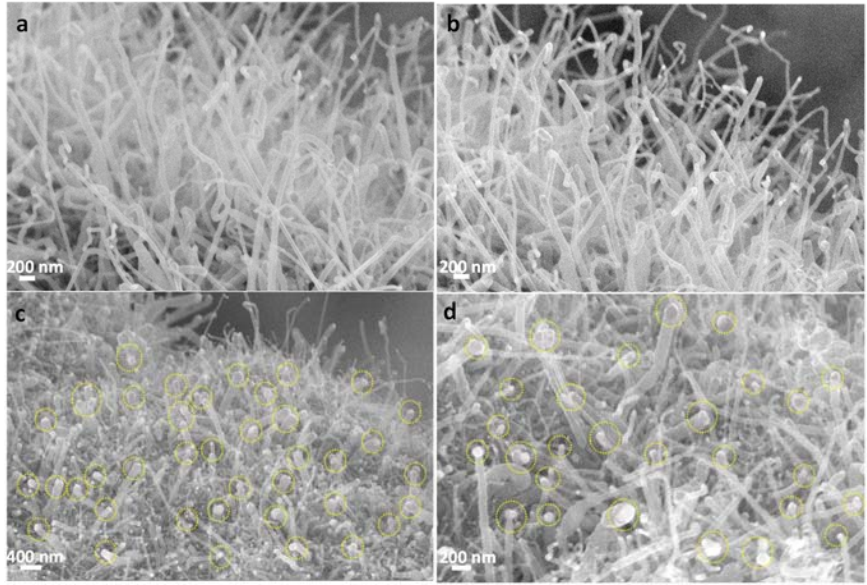

**Fig. S4.** SEM images of (a, b) NCNT/NF and (c, d) CoNi@NCNT/NF. The yellow circle areas denote the carbon-encapsulated CoNi NPs within the top of NCNT arrays.

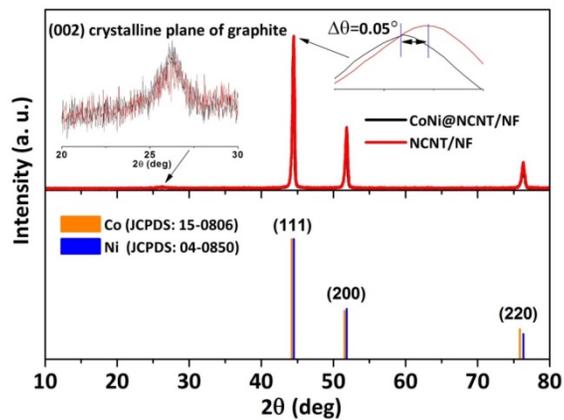

**Fig. S5.** XRD patterns of CoNi@NCNT/NF and NCNT/NF. The inset shows the (002) crystal plane of NCNT at 26°, indicating carbon nanotube arrays were successfully grown on the nickel foam skeleton.

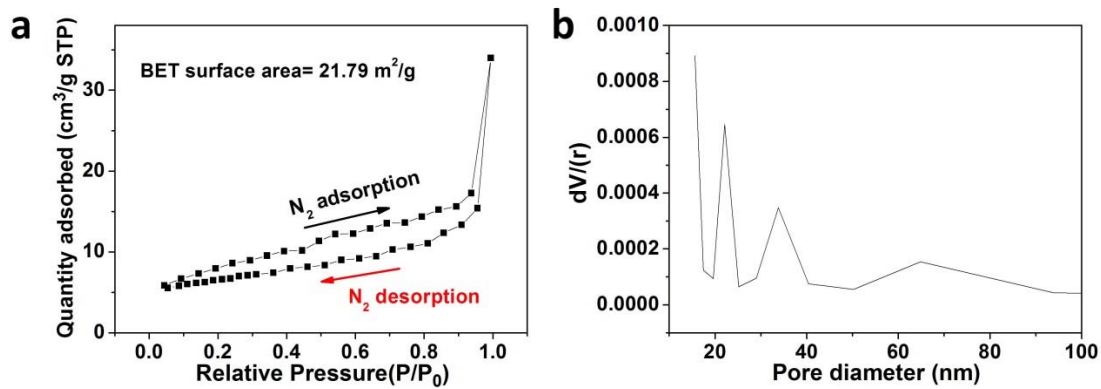

**Fig. S6.** (a) BET surface area and (b) pore size distribution of CoNi@NCNT/NF. Brunauer-Emmett-Teller (BET) analyses were conducted to investigate the specific surface area and pore size distribution of CoNi@NCNT/NF. Nitrogen adsorption-desorption isotherms of CoNi@NCNT/NF (Fig. S6a) show an obvious nitrogen adsorption at the low relative pressure and a hysteresis loops with a wide relative pressure range (0.4-0.9), indicating the co-existence of mesopores and macropores in NCNT arrays. Barrett-Joyner-Halenda (BJH) pore size distribution curves (Fig. S6b) derived from the $N_2$ desorption branches confirm the pore size distribution from 10 to 90 nm.

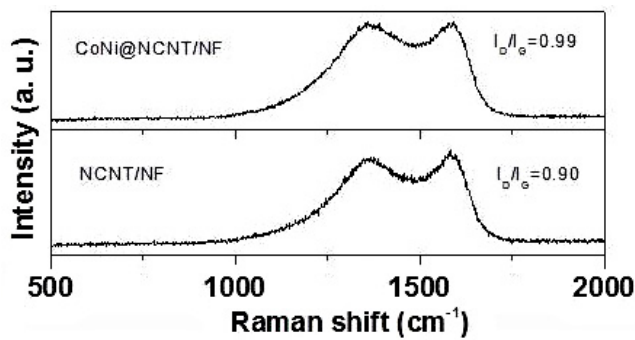

**Fig. S7.** Raman spectrums of CoNi@NCNT/NF, Ni@NCNT/NF and CoNi@NCNT/NF after 28h testing in Zn-air battery.

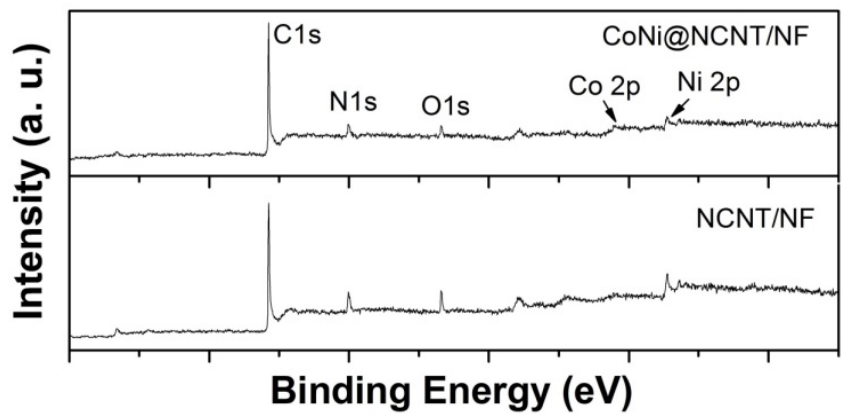

**Fig. S8.** XPS survey spectra of CoNi@NCNT/NF and NCNT/NF.

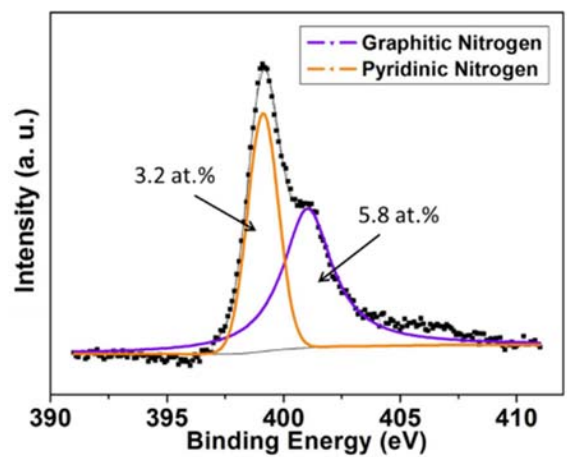

**Fig. S9.** High resolution XPS spectra of N 1s for CoNi@NCNT/NF.

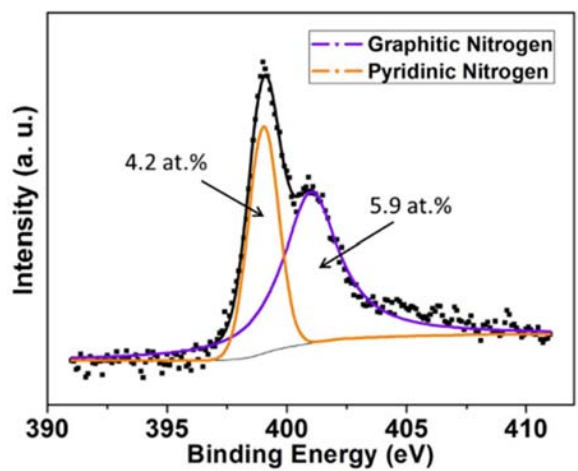

**Fig. S10.** High resolution XPS spectra of N 1s for NCNT/NF.

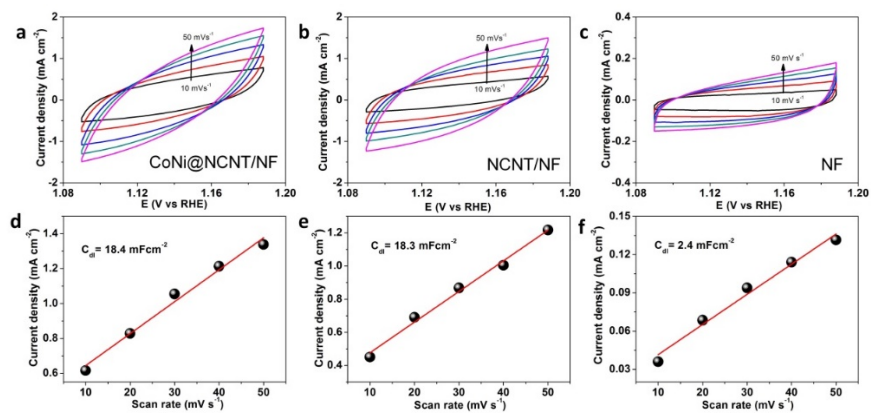

**Supplementary Fig.11.** CVs of (a) CoNi@NCNT/NF, (b) NCNT/NF and (c) NF were recorded in the potential range of 1.08-1.18 V *vs.* RHE at various scan rates and the corresponding linear fitting of the capacitive current densities against the scan rates. The results suggest both CoNi@NCNT/NF and NCNT/NF possess a higher electrochemical active area than that of NF, owing to the NCNT arrays grown on nickel foam.

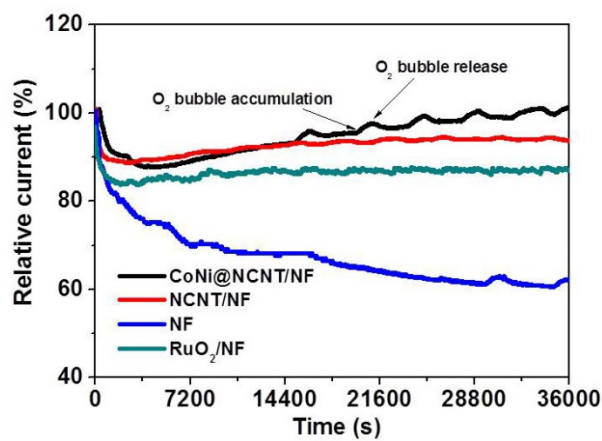

**Fig. S12.** Amperometric (i-t) curves for CoNi@NCNT/NF, NCNT/NF, NF and $RuO_2$/NF electrodes at 1.6 V *vs.* RHE. As can be seen, the current densities are increased with the time increasing for CoNi@NCNT/NF, NCNT/NF and $RuO_2$/NF electrodes, which suggests the nickel foam substrate modified with CoNi@NCNT arrays, NCNT arrays and $RuO_2$ could effective alleviate the loss of active sites, even facilitate the formation of NiOOH with high OER activity. Whereas the NF electrode is directly contacted with electrolyte and $O_2$, which will lead to the slow dissolution of surface active sites on NF.

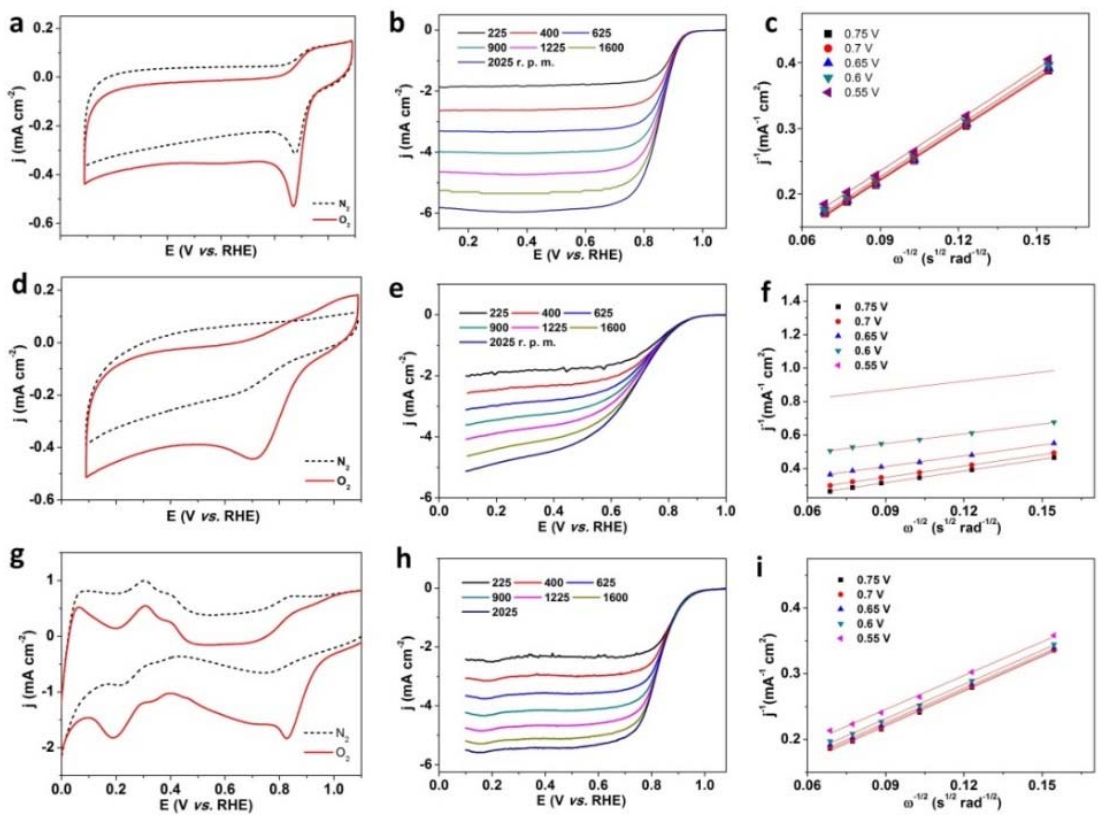

**Fig. S13.** CVs, LSV curves and K-L plots for CoNi@NCNT, NCNT and Pt/C samples in $N_2$ and $O_2$ saturated 0.1 M KOH. The scan rate for CV is 50 mV s$^{-1}$ and for LSV is 5 mV s$^{-1}$.

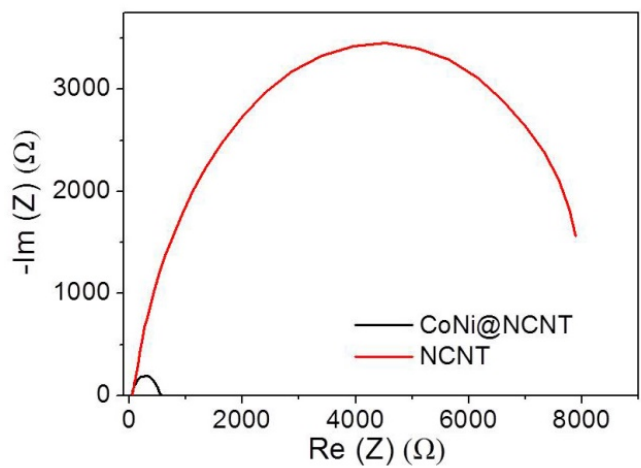

**Supplementary Fig.14.** Electrochemical impedance spectra of CoNi@NCNT and NCNT recorded in $O_2$-saturated 0.1 M KOH at 0.85 V *vs.* RHE with AC amplitude 5 mV and frequency range 10 kHz to 0.01 Hz. The electrode rotation speed is 1600 rpm.

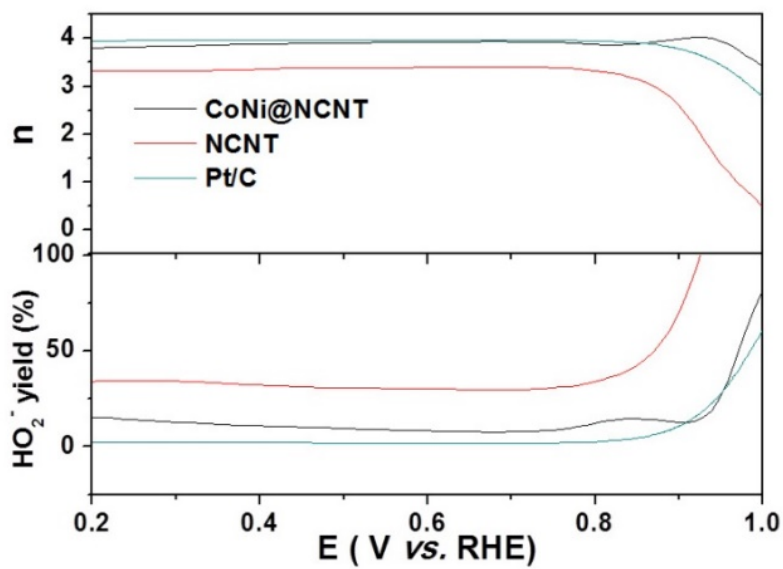

**Fig. S15.** Electron transfer number (n) and $HO_2^-$ yield ratio of all samples.

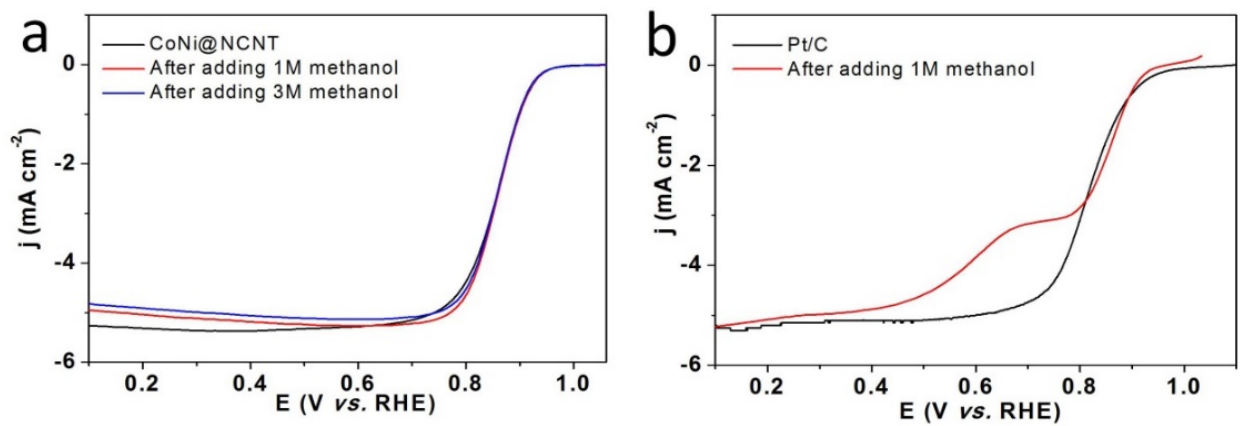

**Fig. S16.** (a) LSVs of CoNi@NCNT electrode with 1600 r. p. m. before and after adding 1 M or 3 M methanol into $O_2$-saturated 0.1 M KOH. (b) LSVs of Pt/C with 1600 r. p. m. before and after adding 1 M methanol into $O_2$-saturated 0.1 M KOH.

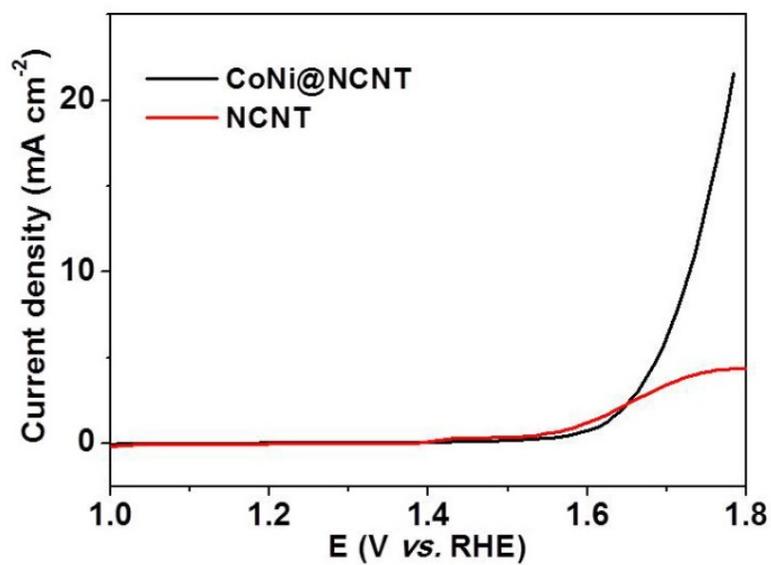

**Fig. S17.** LSVs of CoNi@NCNT and NCNT electrodes in $O_2$-saturated 0.1 M KOH at 1600 rpm.

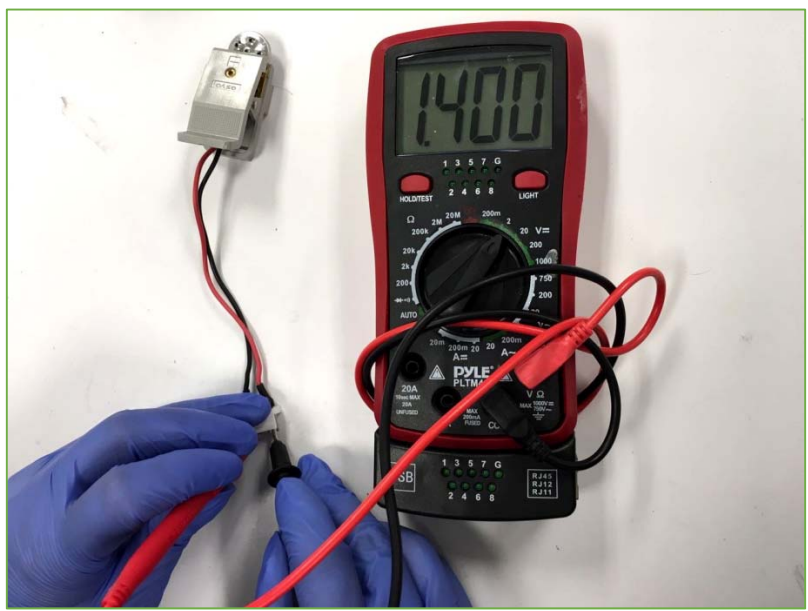

**Fig. S18.** Optical image of an open circuit voltage with 1.4 V for the CoNi@NCNT/NF assembled in a rechargeable Zn-air button battery.

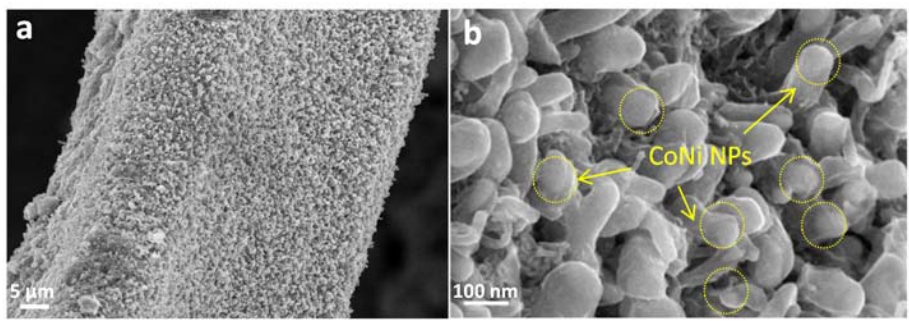

**Fig. S19.** SEM images of the CoNi@NCNT/NF electrode after the long-term galvanostatic charge-discharge cycling over 28h.

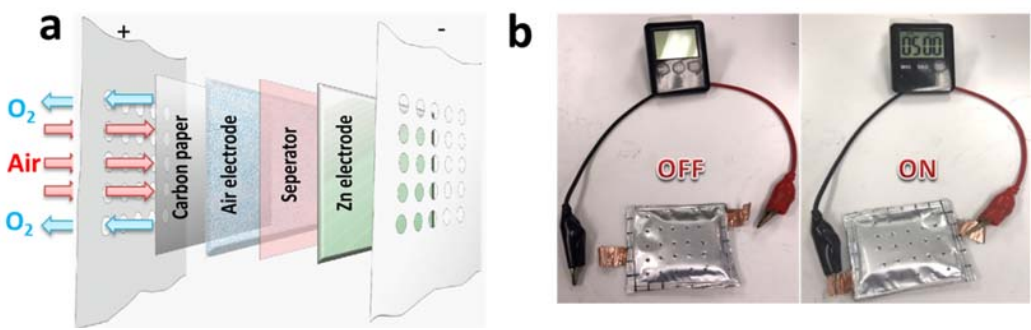

**Fig. S20.** (a) Schematic illustration of a rechargeable Zn-air pouch cell using CoNi@NCNT/NF electrode. (b) Optical images of a timer (1.5 V) being driving by the rechargeable Zn-air pouch cell using CoNi@CNT/NF as cathode.

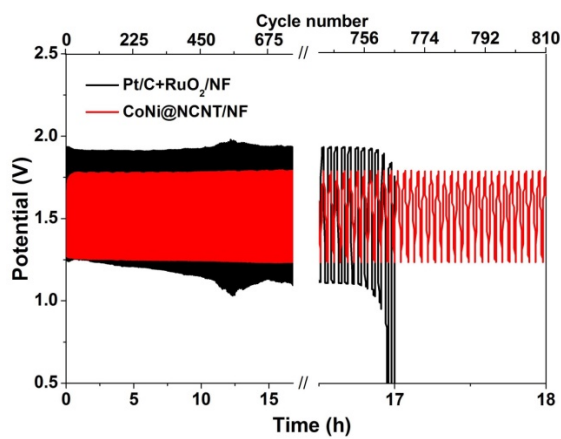

**Fig. S21.** Galvanostatic charge-discharge cycling (2 mA cm$^{-2}$, 80s for each cycle) of the rechargeable Zn-air pouch cell using CoNi@CNT/NF cathode compared with the battery using Pt/C+RuO$_2$/NF air electrode.

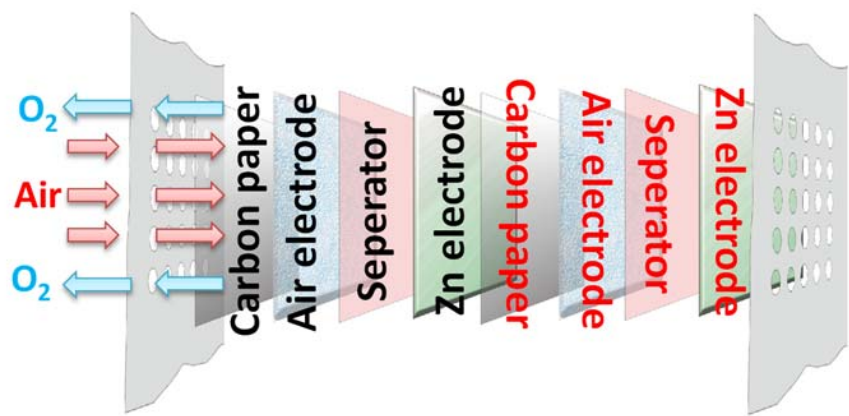

**Fig. S22.** Schematic illustration of two tandem Zn-air components in pouch cell using CoNi@NCNT/NF electrode.

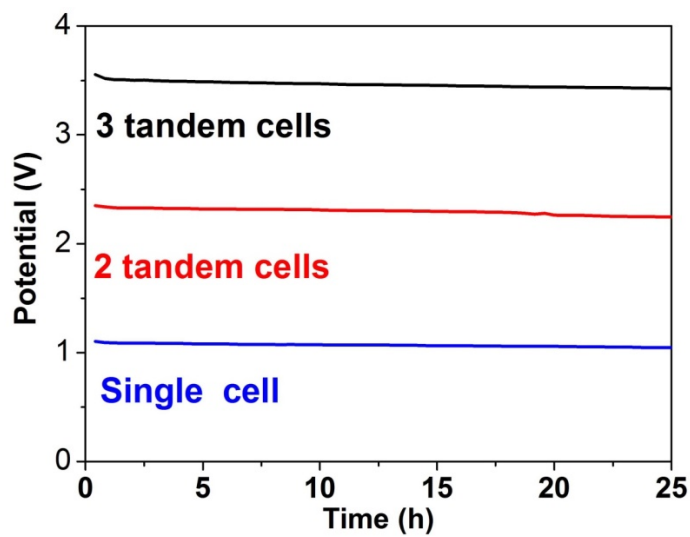

**Fig. S23.** Discharge curves of the pouch cell Zn-air battery with one single, two tandem and three tandem Zn-air components using CoNi@NCNT/NF cathode (discharge current: 10 mA cm$^{-2}$).